\documentclass{article}
\usepackage[english]{babel}
\usepackage[letterpaper,top=2cm,bottom=2cm,left=3cm,right=3cm,marginparwidth=1.75cm]{geometry}
\usepackage{amsmath}
\usepackage{bm}
\usepackage{tabularx}
\newcolumntype{C}{>{\centering\arraybackslash}X}

\usepackage{graphicx}
\usepackage[labelformat=empty]{subcaption}  % disable (a),(b) numbering
\usepackage{subcaption}    
\usepackage[colorlinks=true, allcolors=blue]{hyperref}
\usepackage[percent]{overpic}
\usepackage{authblk}

\title{Single Scattering Properties for an Ensemble of Randomly Oriented Convex Polyhedra in Geometrical Optics Regime}
\author[1,2]{Quan Mu \thanks{Corresponding author: mu.quan@foxmail.com} }
\author[1,2]{Ye Zhang}
\affil[1]{MSU-BIT-SMBU Joint Research Center of Applied Mathematics, Shenzhen MSU-BIT University}
\affil[2]{School of Mathematics and Statistics, Beijing Institute of Technology}

\begin{document}
%\begin{sloppypar}

\maketitle

\begin{abstract}
To study how geometrical shape affect the light scattering properties for an ensemble of randomly orientated particles, the single scattering matrices including complete polarization information are calculated statistically for a group of crystals with random geometrical shape and a group of hexagonal prisms with various aspect ratios in geometrical optics approximation method. To compare, the single scattering matrices for individual random irregular crystal and individual hexagonal prism are also presented. Another important question addressed in this study is whether the regularity of ice crystal shapes can be distinguished from the elements of the scattering matrix. It should be noted that all statistical simulation experiments in this study are restricted to the following conditions: diffraction and absorption effects are neglected, calculations are performed at a single fixed wavelength $0.308 \,\ \mu\text{m}$, the corresponding refractive index of ice is taken as 1.332, particles are assumed to be randomly oriented, and the simulations are limited to the regime where the geometric optics approximation is applicable. Using a unified computational framework for scattering matrices of convex polyhedra, we carried out a series of statistical numerical simulations. The flexibility of this framework in modifying particle geometry enables a systematic investigation of shape-dependent scattering characteristics. The results demonstrate that regular and irregular particles exhibit noticeably different scattering matrix signatures, and ensembles of irregular particles yield smooth and featureless non-zero matrix elements. In contrast, ensembles of regular hexagonal particles with varying aspect ratios retain common geometric scattering features. We further find that ensemble-averaged scattering properties converge with increasing particle number, that elongated needle-like, and plate-like hexagonal prisms do not bound the scattering behavior of hexagonal prisms with intermediate aspect ratios, and that no global linear correlation is observed between aspect ratio and scattering matrix element values. Nevertheless, columnar and plate-like hexagonal crystals can still be distinguished using selected scattering matrix elements at specific scattering angles.

\end{abstract}

\section{Introduction}
Cirrus clouds are generally considered to consist of ice crystals with a wide range of sizes and shapes \cite{liou2016light,mishchenko2000light}. The wide diversity in geometric shape, structural complexity, size scales, and composition poses significant challenges for the characterization of cirrus ice particles, including their microphysical properties. Over the past several decades, considerable efforts have been devoted to improving our understanding of ice cloud particles and their optical scattering characteristics, through ground-based radar and satellite remote sensing \cite{Hemmer2019,acp-9-7115-2009,Mitchell2025acp-25-14071-2025,yang2018review}, in situ aircraft observations \cite{lloyd2020situ,lawson2006situ}, laboratory experiments \cite{schnaiter2016cloud,lamb2023re,MUNOZ20111646}, and numerical approaches \cite{draine1994discrete,mishchenko1996t,yurkin2007discrete}. Among these approaches, numerical calculations of the scattering properties of ice crystals play a fundamental role in interpreting observational data and in developing and optimizing climate models. In particular, the single-scattering matrix provides a complete description of the polarization-dependent scattering behavior of individual particles and serves as a key input for vector radiative transfer simulations and for the development of new parameterization schemes that account for polarization effects. From typical hexagonal prism ice crystals to highly irregular particles, extensive studies have been conducted on their scattering properties and applications in remote sensing and atmospheric modeling. For example, in work \cite{BOROVOI20082648} it is found that polarization of the scattered light is a promising tool for optical diagnostics of crystal shapes. Retrieving aspect ratios of the horizontally oriented hexagonal ice plates from polarization of the scattered light is proposed and discussed. In \cite{YANG20091604}, the dependence of the asymmetry factor on the aspect ratio of randomly oriented hexagonal ice crystals is analyzed.  In \cite{liu2014effective}, it was shown that geometric irregularity and surface roughness are effectively equivalent in determining the single-scattering properties of particles, both producing relatively featureless and smooth scattering phase functions. However, the discussion in that work mainly focuses on only two elements of the scattering matrix: the phase function $M_{11}$ and the degree of linear polarization $-M_{12}/M_{11}$. While extensive calculations of scattering matrices for single particles with various shapes were performed in Ref.~\cite{GRYNKO2003319}, the statistical mean scattering properties of particle ensembles were not addressed. The statistical mean scattering properties of particle ensembles are essential for radiative transfer and remote-sensing applications, since realistic natural media usually consist of large populations of particles with diverse shapes, sizes, and orientations (see, for example, \cite{BARAN20091239,MIN2006161} and references therein). It has been shown that an ensemble model of ice crystals can provide a better representation than single pristine ice crystal models for both the scattered intensity and the polarized intensity of cirrus clouds \cite{Bryan2005,Francis1999}.

In this paper, we calculate single scattering matrices for an ensemble of convex polyhedra to investigate how geometrical shape effect on the non-zero elements of the scattering matrices. By employing the convex hull model and the scattering matrix computation framework introduced in our previous work \cite{Mu2026}, we can calculate single scattering matrices for an ensemble of ice particles with various geometries, even up to thousands, in a statistical sense. This is challenging for most other numerical methods, such as the finite-difference time domain (FDTD) method, the T-matrix method, the discrete dipole approximation (DDA), since each change in particle geometry requires time-consuming preprocessing and computation for each individual particle.  

In section \ref{section 2} we shortly introduce the convex polyhedra, which will be used in our statistical numerical simulation. Computational results and discussion of scattering matrices by an ensemble of convex polyhedra are presented in section \ref{section 3}. Finally, conclusions and remarks are given in section \ref{section 4}. 

\section{Crystal model}
\label{section 2}
In this work, we use the newly developed convex hull model and the unified scattering matrix computational framework in our previous work \cite{Mu2026} to investigate the single scattering properties of a large ensemble of particles with diverse geometrical shapes. The convex hull of a set of points is the minimal convex set that contains all the points. The problem of constructing convex hull of a finite set of points is a classical problem in computational geometry, with broad applications across many fields \cite{preparata2012computational}. Different algorithms and methods for computing convex hulls in two- or three-dimensional spaces have be studied extensively \cite{de2008computational}. Details on the construction of three-dimensional convex bodies and calculation of single scattering matrix can be found in our previous work \cite{mu2022computer,Mu2026} and the references therein, and will not be repeated here. 

\begin{figure*}[htbp]
\centering
\includegraphics[width=0.8\linewidth]{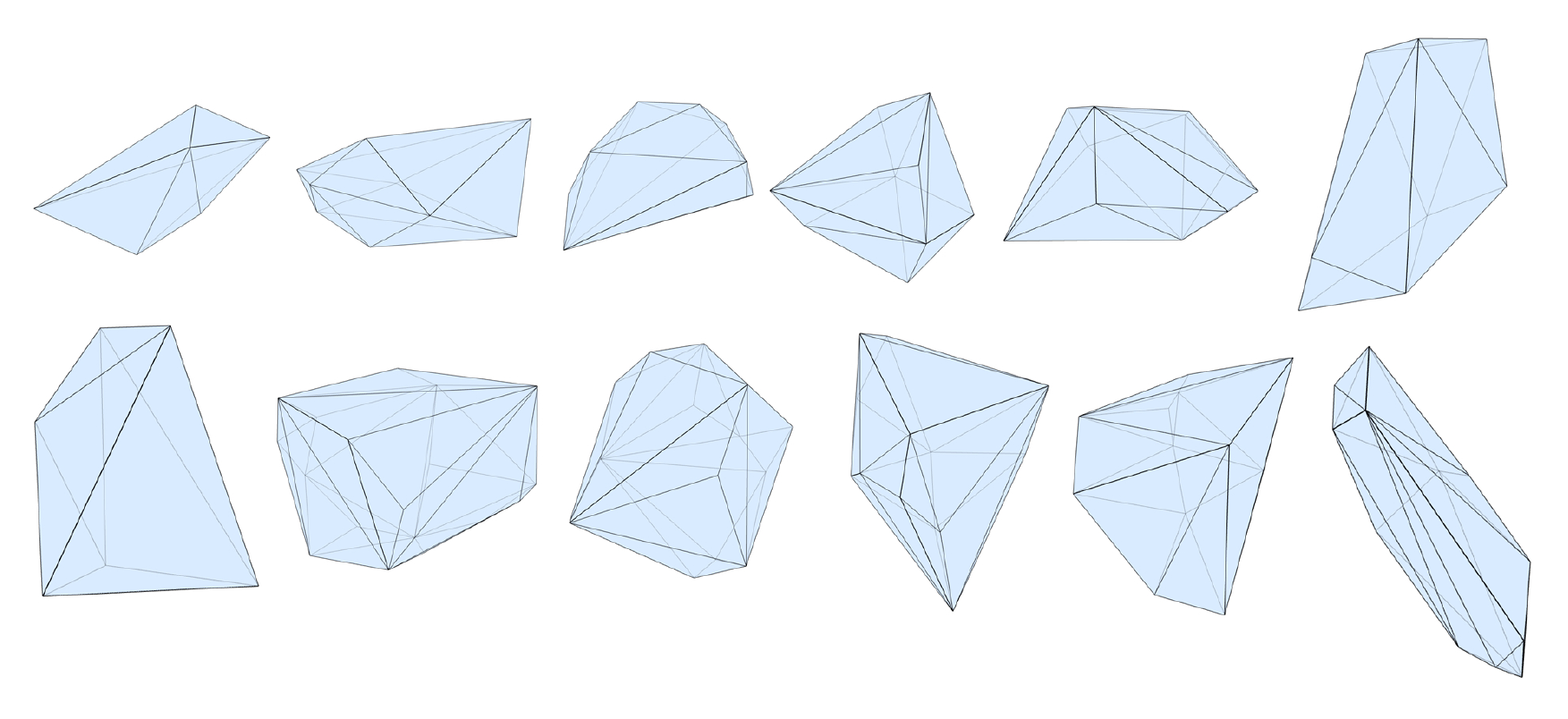}
\caption{\label{Fig1}Examples of random irregular convex hull.}
\end{figure*}

In our previous work \cite{Mu2026} by detecting and merging coplanar triangular faces, the new convex polyhedron model is flexible enough to represent both irregular random convex polyhedra and regular convex polyhedra. Here, we use two relatively simpler models to do the statistical numerical simulation. One model is random convex hull, all the faces of which are triangles. Several examples of this type of random irregular convex hull are illustrated in Figure \ref{Fig1}. Another model is typical hexagonal prism with different aspect ratios. By keeping all other parameters as consistent as possible and varying only the geometry, aspect ratios of ice crystals, and the number of ice particles used in the statistical averaging, we perform a series of statistical numerical simulation experiments. 

To compute the single-scattering matrix of an ensemble of particles in a statistical sense, we first sample a certain number of orientations for each ice crystal. For each orientation, a given number of photons are launched, and each photon is traced through a controlled number of internal reflections and refractions before the simulation is terminated. The resulting Jones matrices are converted into Mueller matrices, and then their contributions are accumulated into bins according to the scattering angle. Finally, after normalizing $M_{11}$, the remaining elements are expressed relative to $M_{11}$.   

\section{Results and discussion}
\label{section 3}
We first performed a series of numerical statistical computations of single-scattering matrices for typical hexagonal prisms, followed by those for random convex hulls. It should be noted that all statistical simulation experiments in this study are conducted under the following assumptions: diffraction and absorption effects are neglected; calculations are performed at a single fixed wavelength of $0.308 \mu\text{m}$; the refractive index of ice is set to 1.332; particles are assumed to be randomly oriented; and the simulations are restricted to the validity regime of the geometric optics approximation. The corresponding results and discussions are provided in the following subsections. 

\subsection{Hexagonal prism}
\begin{figure*}[htbp]
    \centering
    % Panel B
    \begin{subfigure}[b]{0.18\textwidth}
        \centering
        \includegraphics[width=\textwidth]{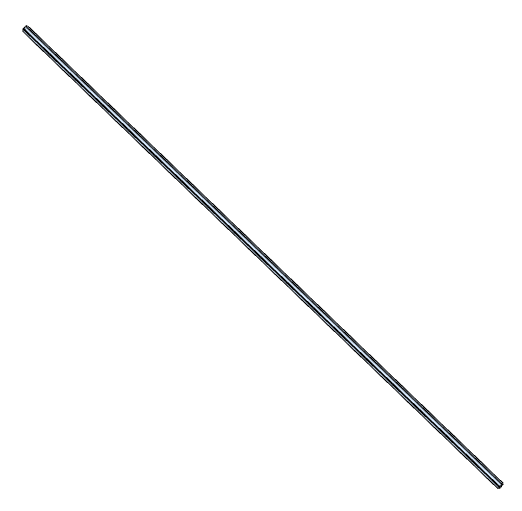}
        \caption{B}
    \end{subfigure}
    \hfill
    % Panel C
    \begin{subfigure}[b]{0.18\textwidth}
        \centering
        \includegraphics[width=\textwidth]{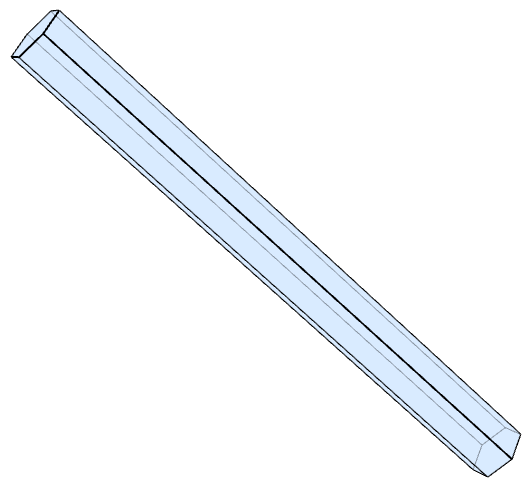}
        \caption{C}
    \end{subfigure}
    \hfill
    % Panel D
    \begin{subfigure}[b]{0.18\textwidth}
        \centering
        \includegraphics[width=\textwidth]{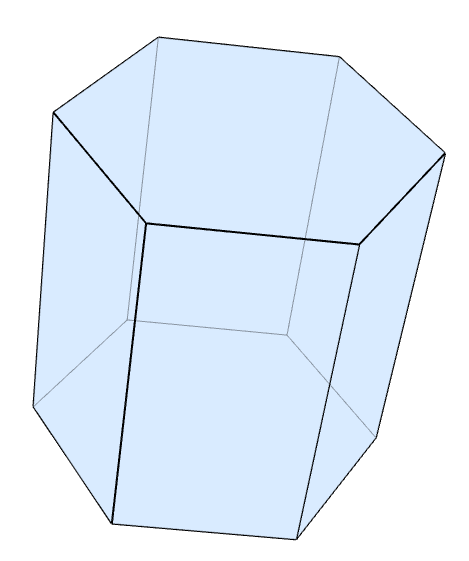}
        \caption{D}
    \end{subfigure}
    \hfill
    % Panel E
    \begin{subfigure}[b]{0.18\textwidth}
        \centering
        \includegraphics[width=\textwidth]{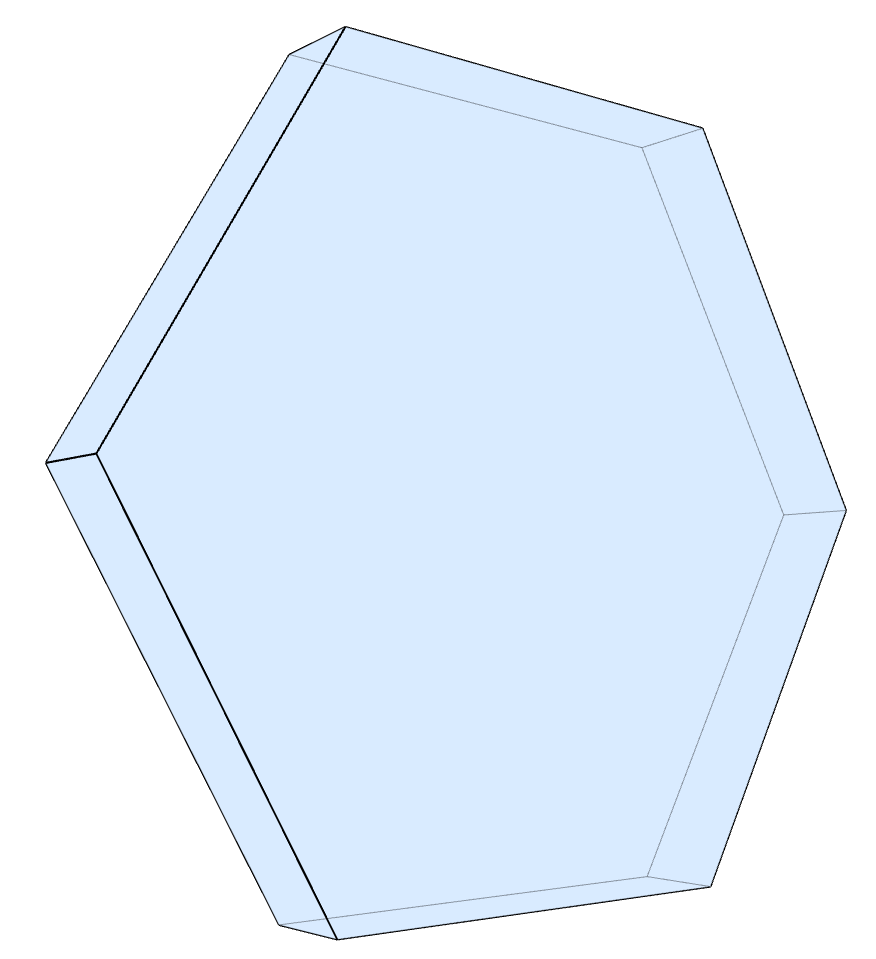}
        \caption{E}
    \end{subfigure}
    \hfill
    % Panel F
    \begin{subfigure}[b]{0.18\textwidth}
        \centering
        \includegraphics[width=\textwidth]{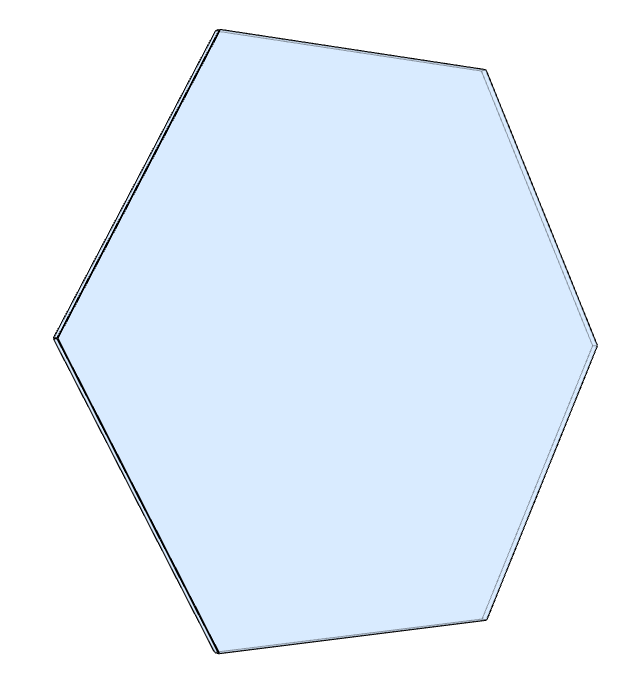}
        \caption{F}
    \end{subfigure}

    \caption{Geometric representation of 5 hexagonal prisms with aspect ratios $r = 0.01, 0.1, 1, 10, 100$, corresponding to panels B, C, D, E, F, respectively.}
    \label{Fig2}
\end{figure*}

\begin{figure*}[htbp]
\centering

\begin{overpic}[width=0.48\textwidth]{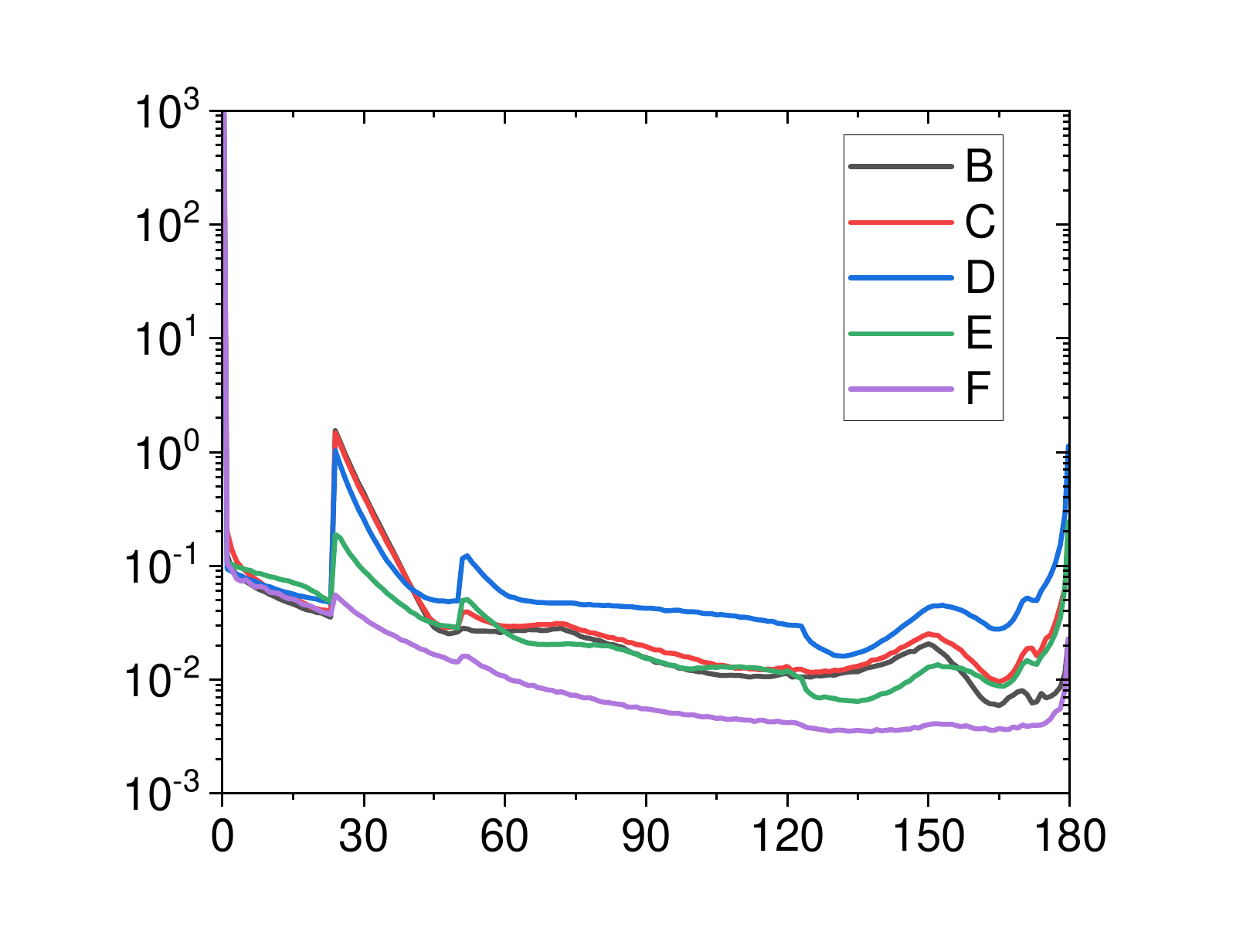}
  \put(20,15){\small $M_{11}$}
\end{overpic}
\begin{overpic}[width=0.48\textwidth]{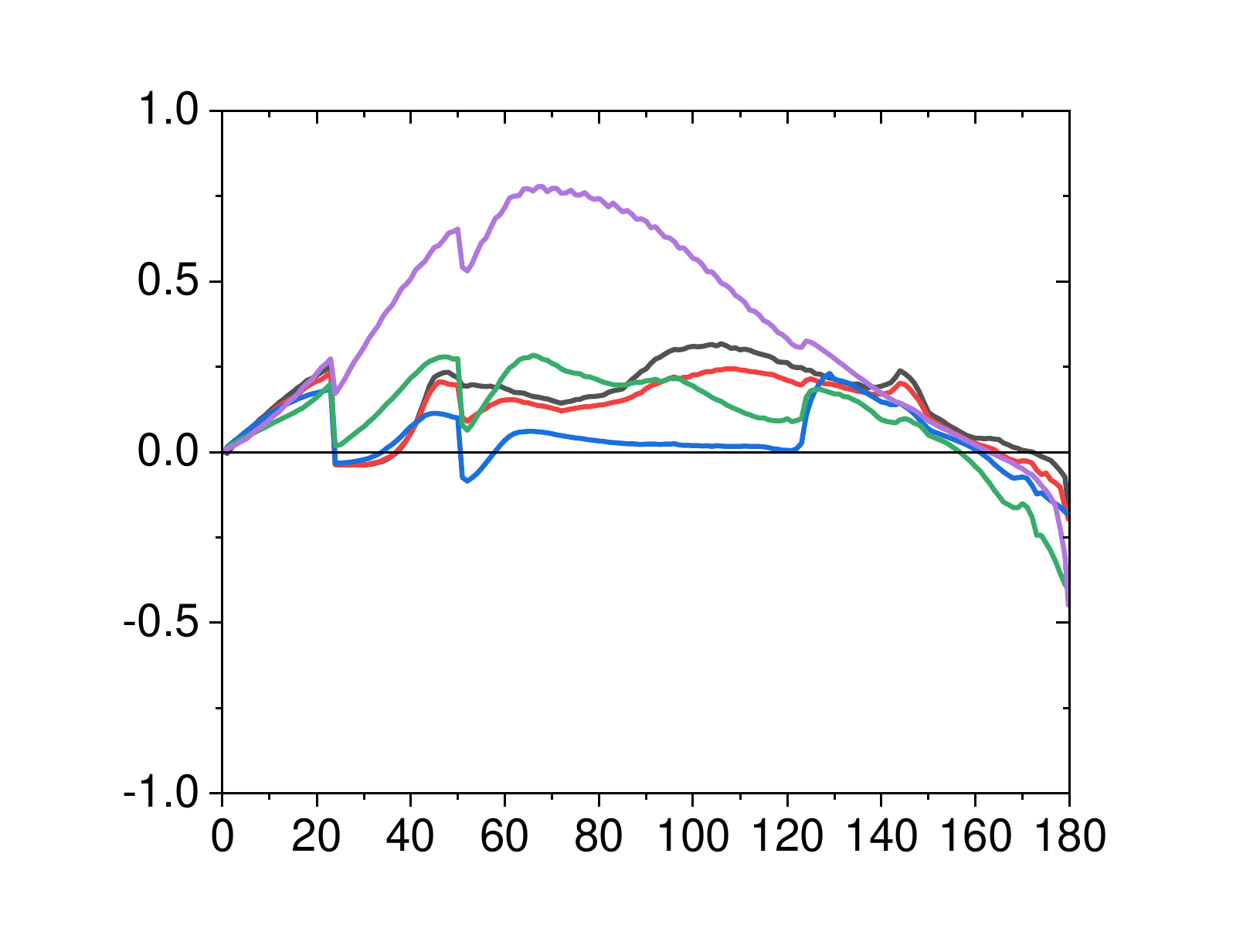}
  \put(20,15){\small $M_{12}/M_{11}$}
\end{overpic}
%---------------------------------------------
\begin{overpic}[width=0.48\textwidth]{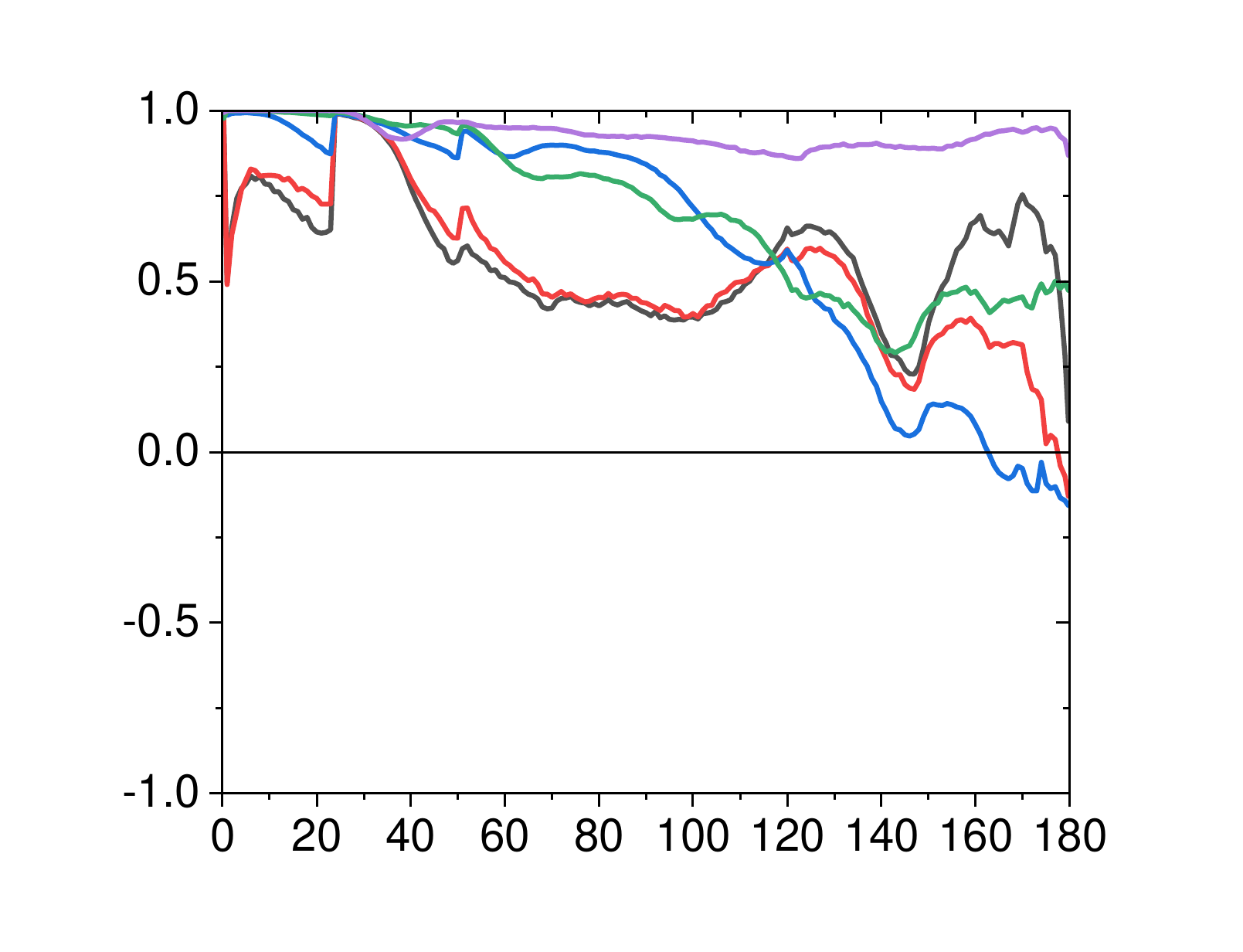}
  \put(20,15){\small $M_{22}/M_{11}$}
\end{overpic}
\begin{overpic}[width=0.48\textwidth]{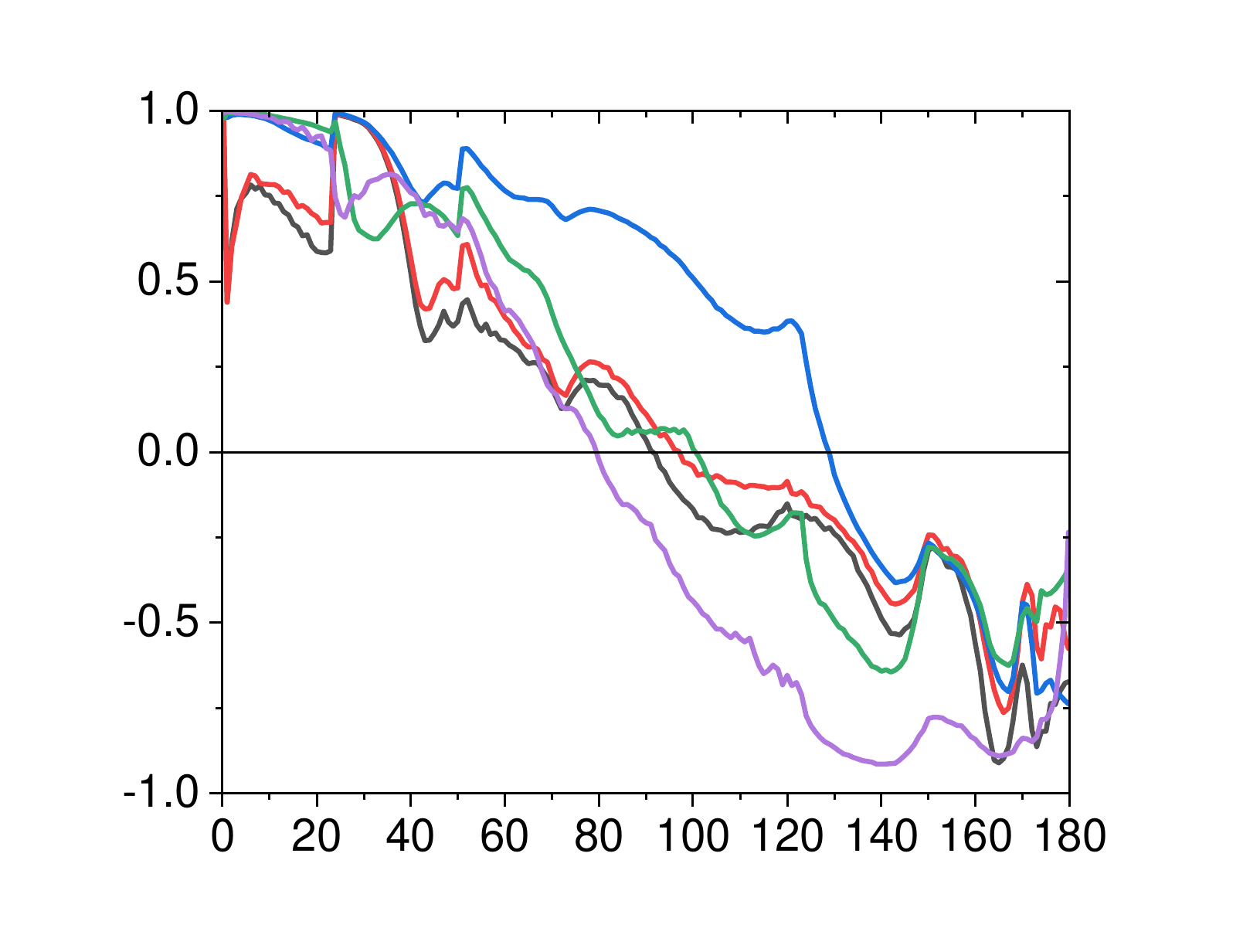}
  \put(20,15){\small $M_{33}/M_{11}$}
\end{overpic}
%---------------------------------------------
\begin{overpic}[width=0.48\textwidth]{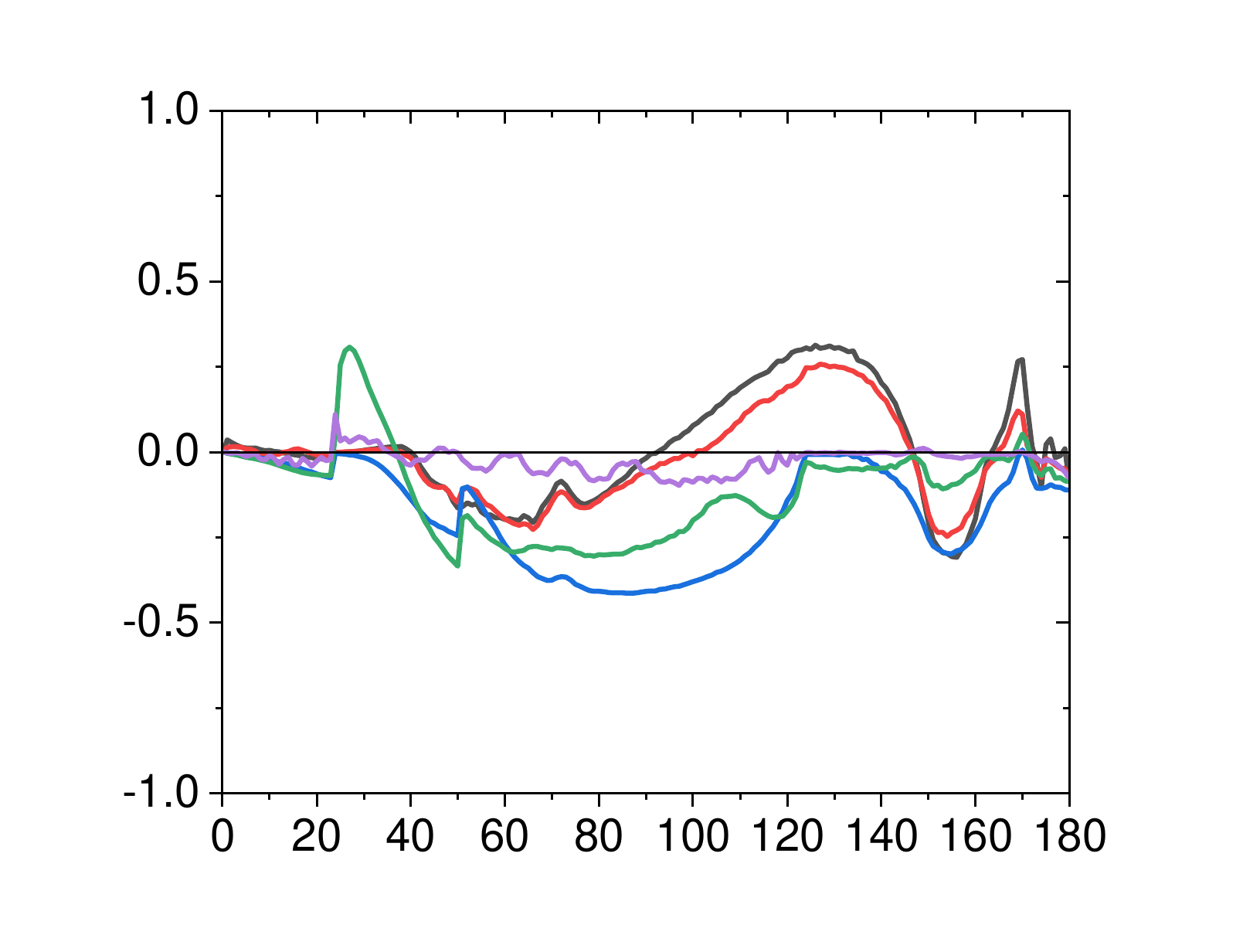}
  \put(20,15){\small $M_{34}/M_{11}$}
\end{overpic}
\begin{overpic}[width=0.48\textwidth]{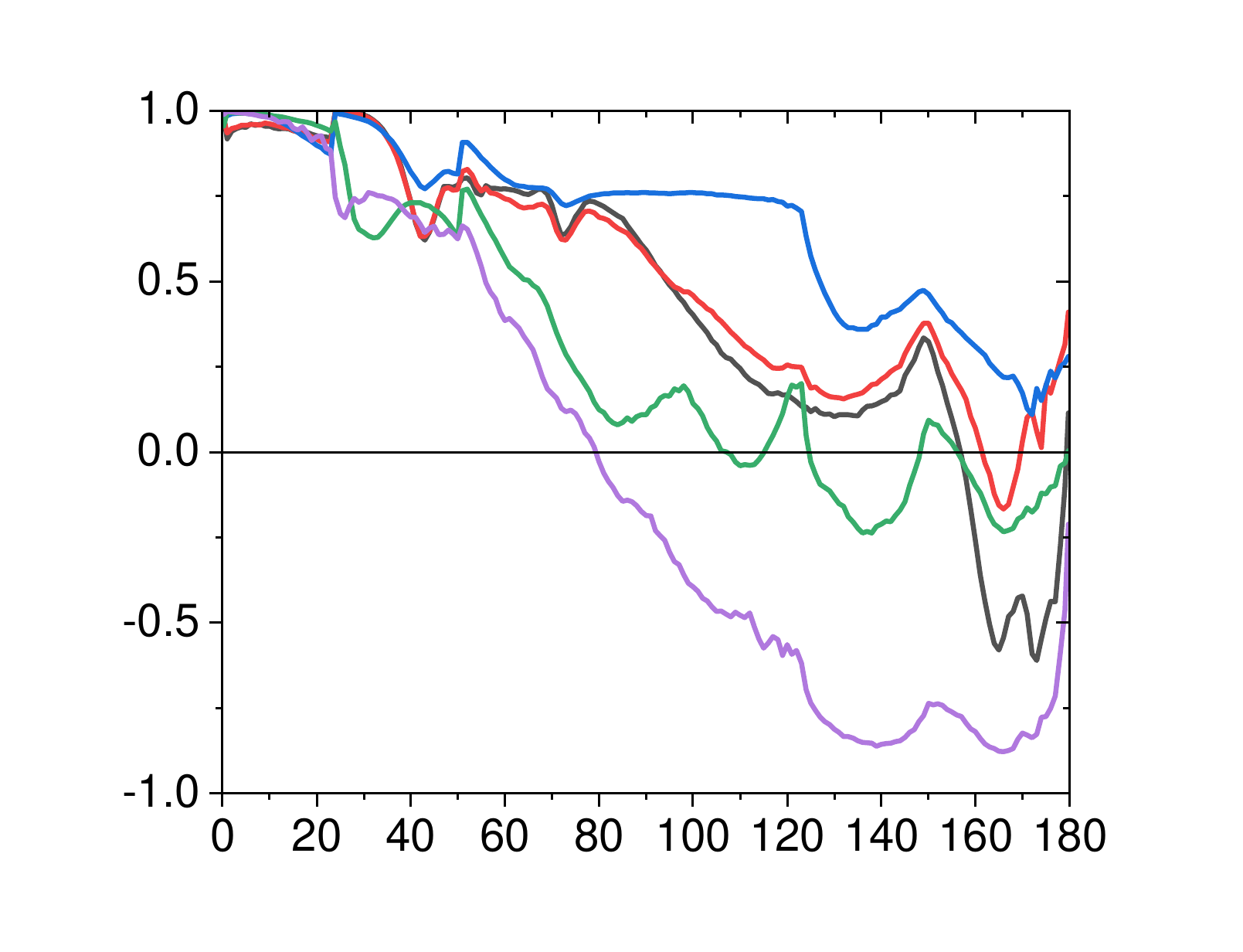}
  \put(20,15){\small $M_{44}/M_{11}$}
\end{overpic}

\caption{Scattering matrix elements for 5 individual randomly oriented hexagonal prisms with aspect ratios $r = 0.01, 0.1, 1, 10, 100$, corresponding to panels B, C, D, E, F in Figure~\ref{Fig2} respectively. The horizontal axis represents the scattering angle (in degrees).}
\label{Fig3}
\end{figure*}

The scattering matrices were computed for five individual, randomly oriented hexagonal prisms with different aspect ratios $r = d/h$, where $d$ is the base diameter and $h$ is the prism height. To investigate their scattering properties, hexagonal prisms with aspect ratios $r = 0.01, 0.1, 1, 10,$ and $100$ were considered, ranging from extremely elongated needle-like to ultra-thin plate-like forms. The corresponding particle shapes are shown in panels B--F of Figure~\ref{Fig2}, respectively. Since absorption is neglected, the specific physical scale does not affect the results. The number of traced rays is set to $100$ for each orientation, and the number of sampled orientation is set to $10^6$ for $r = 0.1, 1, 10$. Since the unified framework employs a Monte Carlo \textit{hit-and-miss} approach, a larger number of particle orientations are required for extremely elongated columnar crystals and ultra-thin plate-like crystals in order to capture sufficient valid ray paths. Therefore, for $r = 0.01$ and $r = 100$, the number of sampled orientations is set to $5 \times 10^{6}$ and $3 \times 10^{6}$, respectively. The maximum number of internal reflections inside the scatterer is set to $10$. The calculated scattering matrices are presented in Figure~\ref{Fig3}.
%%--------------------------------------------------------------Fig2

%%--------------------------------------------------------------Fig3

To explore the influence of particle geometry, we deliberately included extremely elongated columnar crystals and ultra-thin plate-like crystals in the simulations. The aim was to investigate whether these two limiting shapes could serve as boundaries for the scattering characteristics of hexagonal crystals with intermediate aspect ratios, that is, whether the scattering matrix element values of other particles are constrained between these two extremes. Moreover, we aimed to determine whether linear correlations between specific scattering matrix elements and the aspect ratio $r$ could be identified at particular scattering angles. As shown in Figure~\ref{Fig3}, no clear global limiting behavior is observed for the scattering matrix elements over the full angular range. Linear correlations between scattering matrix elements and the aspect ratio $r$ are only found within very limited angular regions. For example, they are observed in $M_{11}$ around $30^\circ$--$40^\circ$ and in $M_{22}/M_{11}$ around $45^\circ$--$50^\circ$. It should be noted that these findings are based on a limited dataset, and further numerical tests are essential to assess their robustness.

Interestingly, we find that the curves corresponding to the blue ($r = 1$) and purple ($r = 100$) cases tend to form envelope-like bounds across several angular regions in most scattering matrix elements. Also, it may be possible to distinguish columnar and plate-like geometries based on certain scattering matrix elements in selected angular intervals. Furthermore, within the respective bounds of columnar and plate-like crystals, linear correlations between certain scattering matrix elements and the aspect ratio $r$ may exist at specific scattering angles. Evidence supporting these conclusions and hypotheses can be found in several regions in Figure~\ref{Fig3}, for example near $40^\circ$ in $M_{22}/M_{11}$, around $130^\circ$ in $M_{34}/M_{11}$, near $140^\circ$ in $M_{44}/M_{11}$, and near the backward scattering direction ($180^\circ$) in $M_{22}/M_{11}, M_{44}/M_{11}$.

From the $M_{11}$ element in Figure~\ref{Fig3}, we can clearly observe several classical scattering features of hexagonal ice crystals across different aspect ratios $r$, including the pronounced delta transmission peak in the forward direction and well-known halo peaks at $22^\circ$ and $46^\circ$. It is worth noting that for extremely elongated needle-like ($r = 0.01$) and ultra-thin plate-like ($r = 100$) ice crystals, the intensity enhancement at $46^\circ$ becomes very weak. This behavior can be attributed to the fact that, for these two extreme geometries, the probability of light rays following the optical paths responsible for the formation of the $46^\circ$ halo is significantly reduced (for more detailed explanations of the ray paths responsible for the formation of specific halos, see, e.g., Ref.~\cite{tape2006atmospheric, tape2013atmospheric}). Polarization signatures associated with these halo angles are also evident in other scattering matrix elements. In particular, polarization information at the halo scattering directions can be identified in $M_{12}/M_{11}$, $M_{22}/M_{11}$, and $M_{34}/M_{11}$, reflecting pronounced linear polarization features and the presence of linear--circular polarization coupling at scattering angles near $22^\circ$ and $46^\circ$.

\begin{figure*}[htbp]
\centering

\begin{overpic}[width=0.48\textwidth]{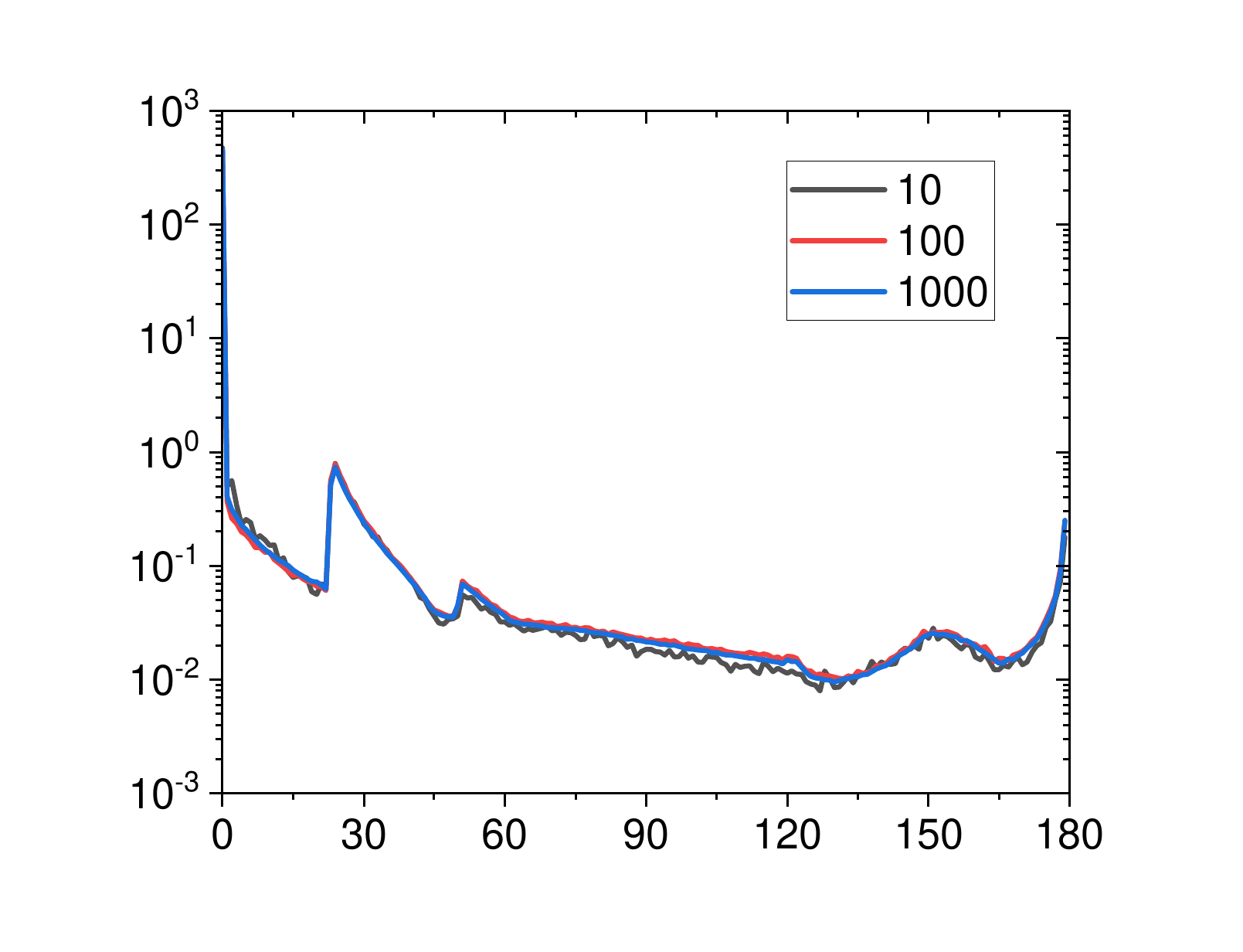}
  \put(20,15){\small $M_{11}$}
\end{overpic}
\begin{overpic}[width=0.48\textwidth]{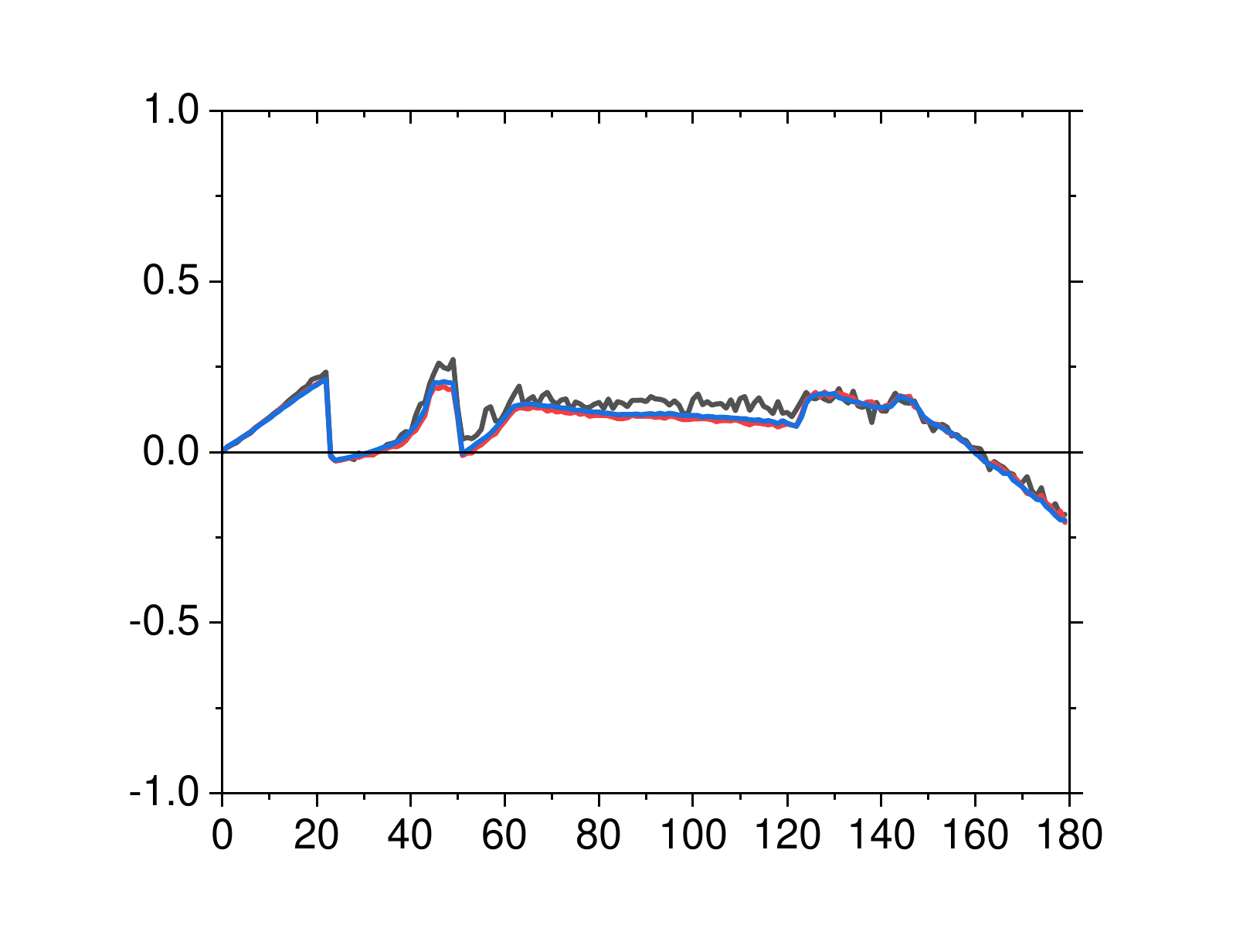}
  \put(20,15){\small $M_{12}/M_{11}$}
\end{overpic}
%---------------------------------------------
\begin{overpic}[width=0.48\textwidth]{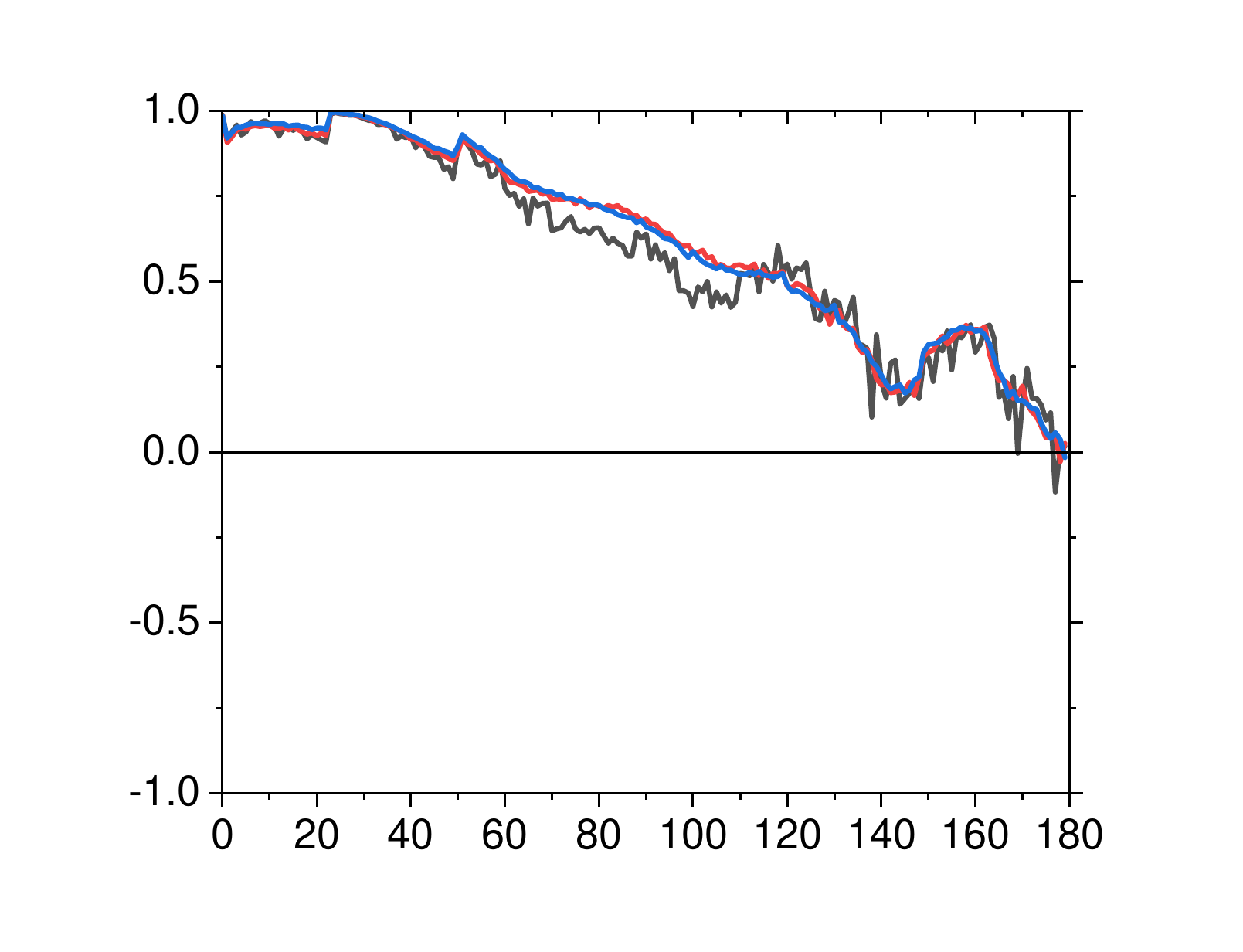}
  \put(20,15){\small $M_{22}/M_{11}$}
\end{overpic}
\begin{overpic}[width=0.48\textwidth]{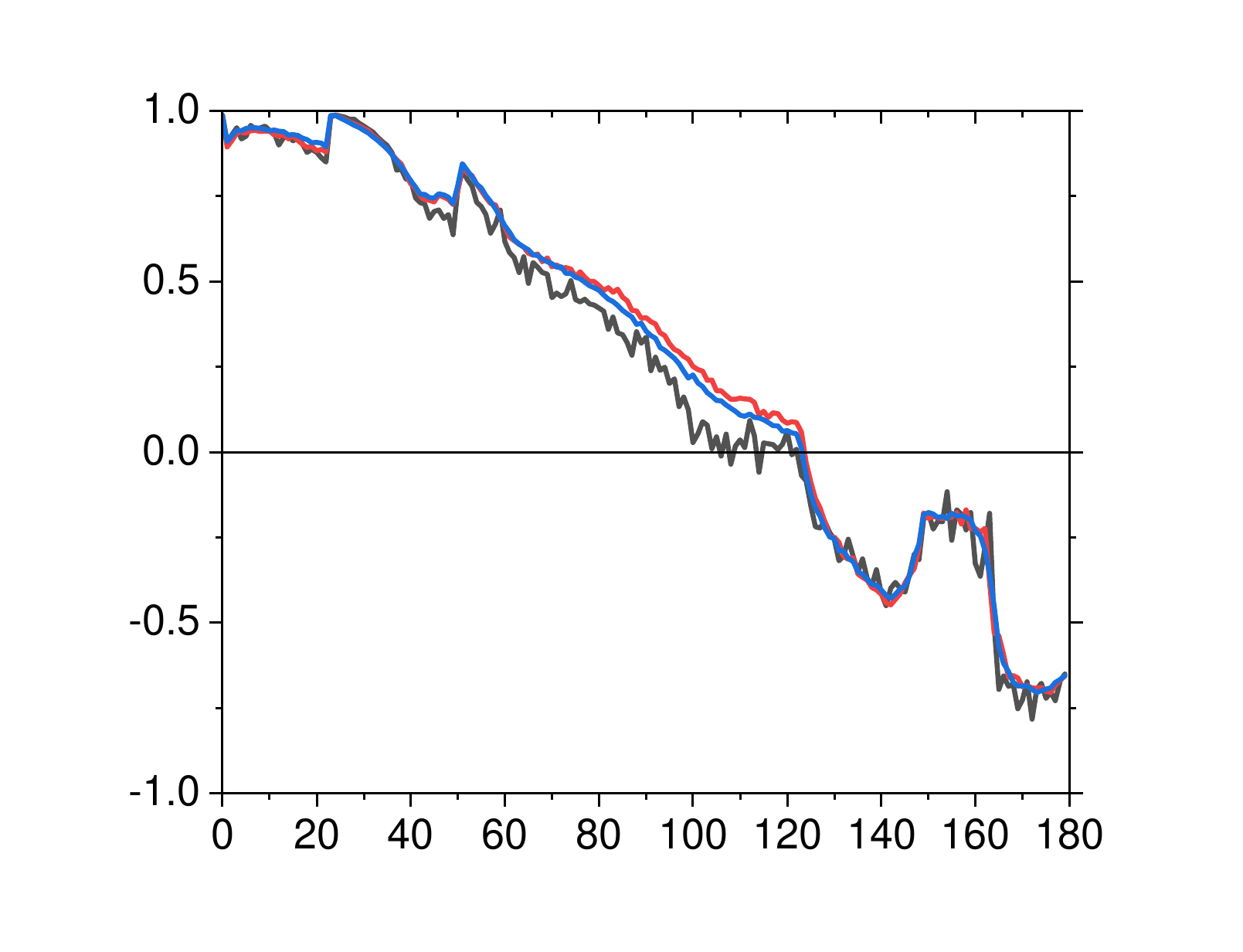}
  \put(20,15){\small $M_{33}/M_{11}$}
\end{overpic}
%---------------------------------------------
\begin{overpic}[width=0.48\textwidth]{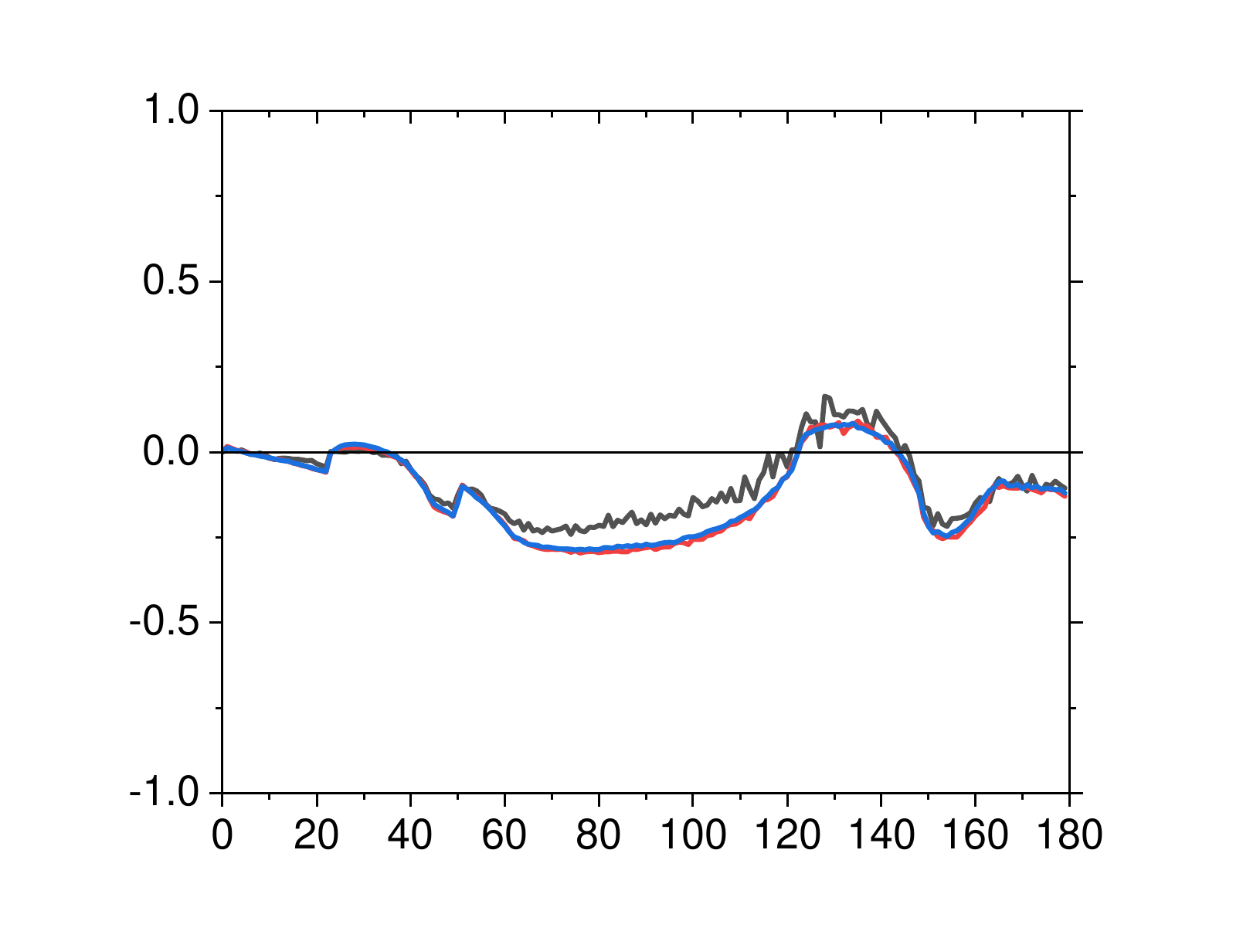}
  \put(20,15){\small $M_{34}/M_{11}$}
\end{overpic}
\begin{overpic}[width=0.48\textwidth]{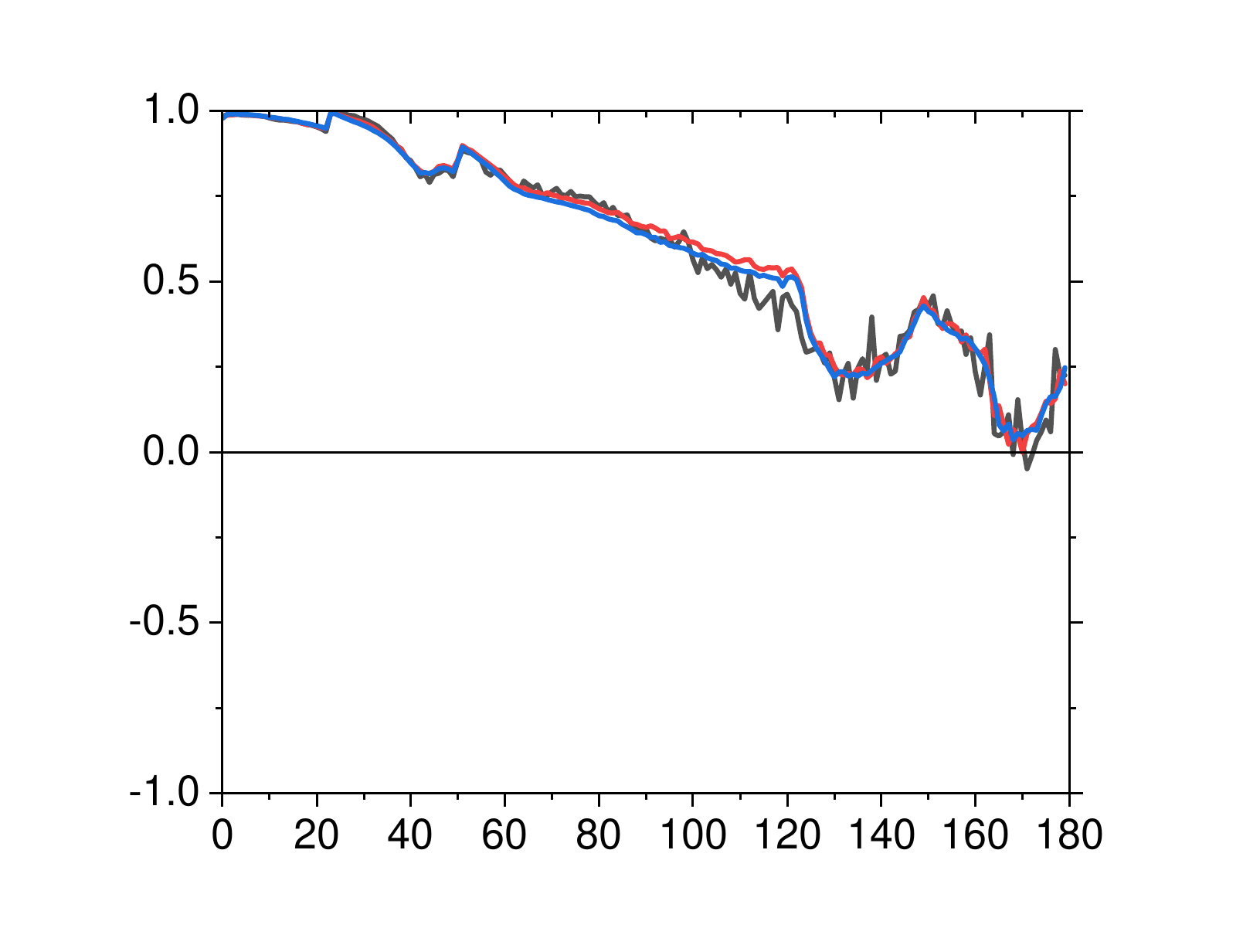}
  \put(20,15){\small $M_{44}/M_{11}$}
\end{overpic}

\caption{Single-scattering matrix elements computed for three ensembles of hexagonal prisms, each consisting of 10, 100, and 1000 particles with randomly sampled aspect ratios. The horizontal axis represents the scattering angle (in degrees).}
\label{Fig6}
\end{figure*}

In Figure~\ref{Fig6}, the six scattering matrix elements are shown for three ensembles of hexagonal prisms, containing 10, 100, and 1000 particles. For each ensemble, the particle aspect ratios $r$ are uniformly sampled from four intervals $[0.01,0.1]$, $[0.1,1]$, $[1,10]$, and $[10,100]$ with corresponding weights of 0.2, 0.3, 0.3, and 0.2, respectively. The maximum number of internal reflections inside the scatterer is set to $5$, the number of sampled orientation for each particle is set to $10^3$, and to ensure that the total number of sampled photons is the same ($10^4$) for different ensemble sizes, the number of traced rays per particle is set to 1000, 100, and 10 for ensemble sizes of 10, 100, and 1000, respectively. The resulting single scattering matrices represent statistical averages over the ensembles. Using the unified computational framework introduced in our previous work \cite{Mu2026}, we are now able to perform those numerical statistical simulations, which are essential for statistical modeling in random scattering medium.

%%--------------------------------------------------------------Fig6

From Figure~\ref{Fig6}, we find that, in general, all nonzero elements of the scattering matrix computed for ensembles with different sizes tend to converge to the same values over all scattering directions. In particular, the scattering matrices obtained for ensembles containing 100 and 1000 particles with different aspect ratios are in close agreement, indicating that further increasing the ensemble size does not lead to noticeable changes in the ensemble-averaged results and that statistical convergence has been essentially achieved. In contrast, the ensemble consisting of only 10 particles exhibits numerous small spiky fluctuations in the scattering matrix elements, which can be mainly attributed to the insufficient number of particles used in the averaging process.

Moreover, it is clearly observed that the ensemble-averaged scattering matrices still retain the characteristic $22^\circ$ and $46^\circ$ halo features, and the forward delta-transmission peak remains pronounced. This behavior arises because, although the particles considered have different aspect ratios, they all preserve the hexagonal crystal structure, so that the interfacial angles between any two faces remain unchanged.

By comparing Figures~\ref{Fig3} and~\ref{Fig6}, it can be observed that the ensemble-averaged scattering matrix elements generally lie within the range spanned by the scattering matrix elements of the individual particles. This behavior primarily arises from the symmetric, weighted sampling presetting adopted for the particle aspect ratios, as previously described in our numerical experiments: for each ensemble, the particle aspect ratios $r$ are uniformly sampled from four intervals $[0.01, 0.1]$, $[0.1, 1]$, $[1, 10]$, and $[10, 100]$, with corresponding weights of 0.2, 0.3, 0.3, and 0.2, respectively.

In practical applications and in the interpretation of remote-sensing observations, the ensemble scattering characteristics can be tailored by adjusting the sampling distributions of aspect ratio and their associated weights to better represent realistic ice crystal populations or meet specific observational requirements. This flexibility is useful for improving climate and atmospheric models and for interpreting the relative contributions and proportions of ice crystals with different aspect ratios in cirrus clouds. The computational framework proposed in our previous work \cite{Mu2026} can be used to address this type of requirement.

\subsection{Random convex hull}

\begin{figure*}[htbp]
    \centering
    % Panel B
    \begin{subfigure}[b]{0.15\textwidth}
        \centering
        \includegraphics[width=\textwidth]{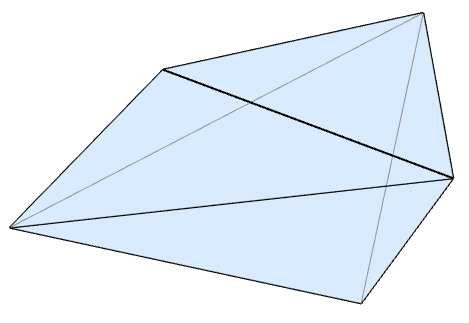}
        \caption{B}
    \end{subfigure}
    \hfill
    % Panel C
    \begin{subfigure}[b]{0.15\textwidth}
        \centering
        \includegraphics[width=\textwidth]{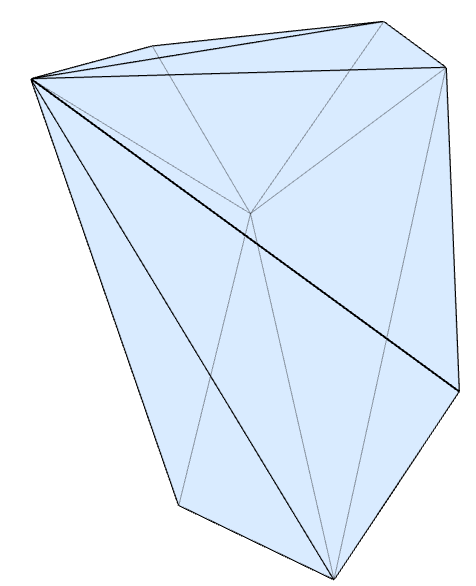}
        \caption{C}
    \end{subfigure}
    \hfill
    % Panel D
    \begin{subfigure}[b]{0.15\textwidth}
        \centering
        \includegraphics[width=\textwidth]{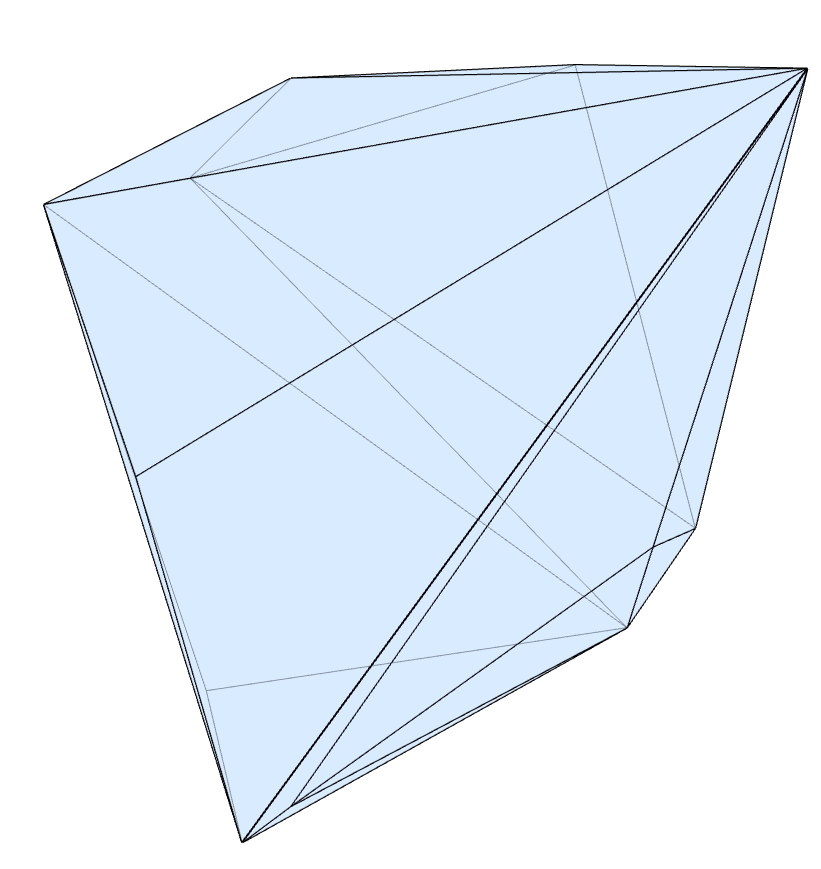}
        \caption{D}
    \end{subfigure}
    \hfill
    % Panel E
    \begin{subfigure}[b]{0.15\textwidth}
        \centering
        \includegraphics[width=\textwidth]{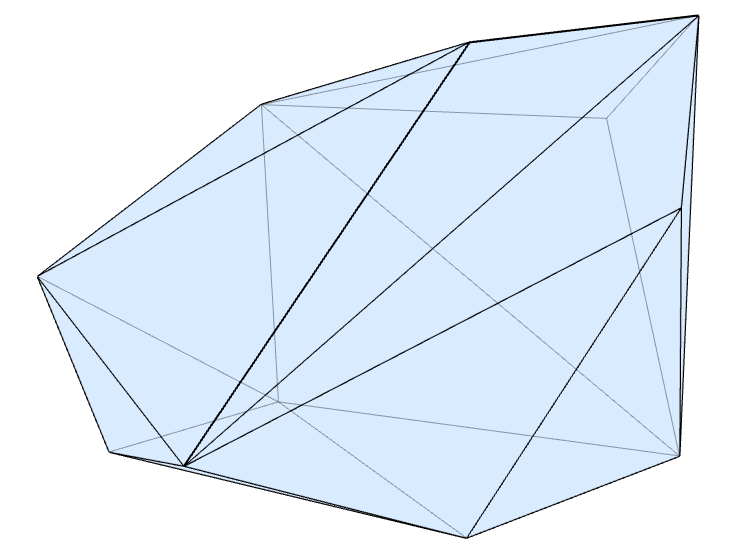}
        \caption{E}
    \end{subfigure}
    \hfill
    % Panel F
    \begin{subfigure}[b]{0.15\textwidth}
        \centering
        \includegraphics[width=\textwidth]{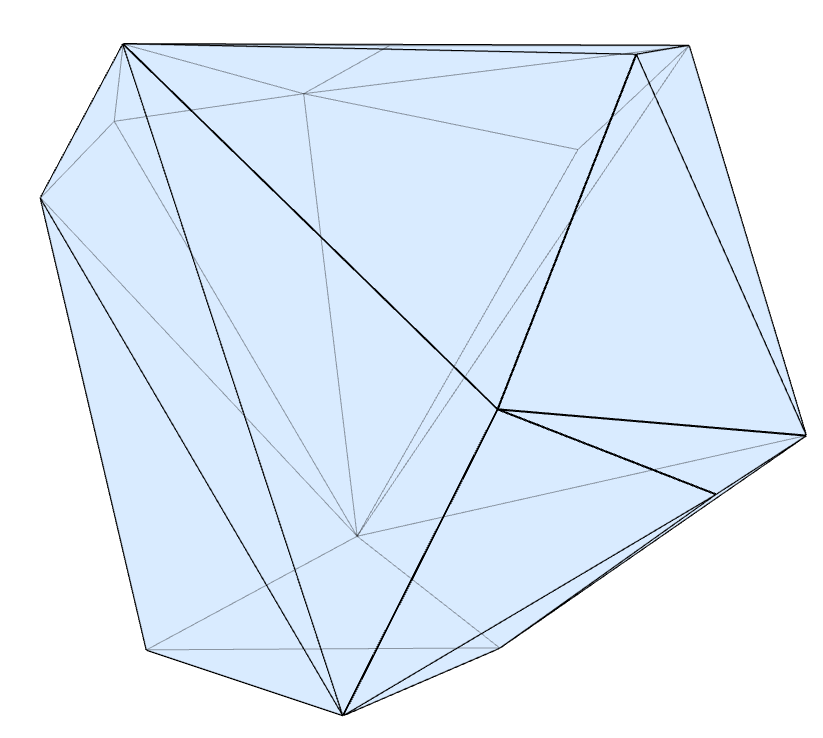}
        \caption{F}
    \end{subfigure}

    \caption{Geometric representation of 5 randomly generated convex hulls with NIP = 6, 8, 13, 18, 30, corresponding to panels B, C, D, E, F, respectively.}
    \label{Fig4}
\end{figure*}

\begin{figure*}[htbp]
\centering

\begin{overpic}[width=0.48\textwidth]{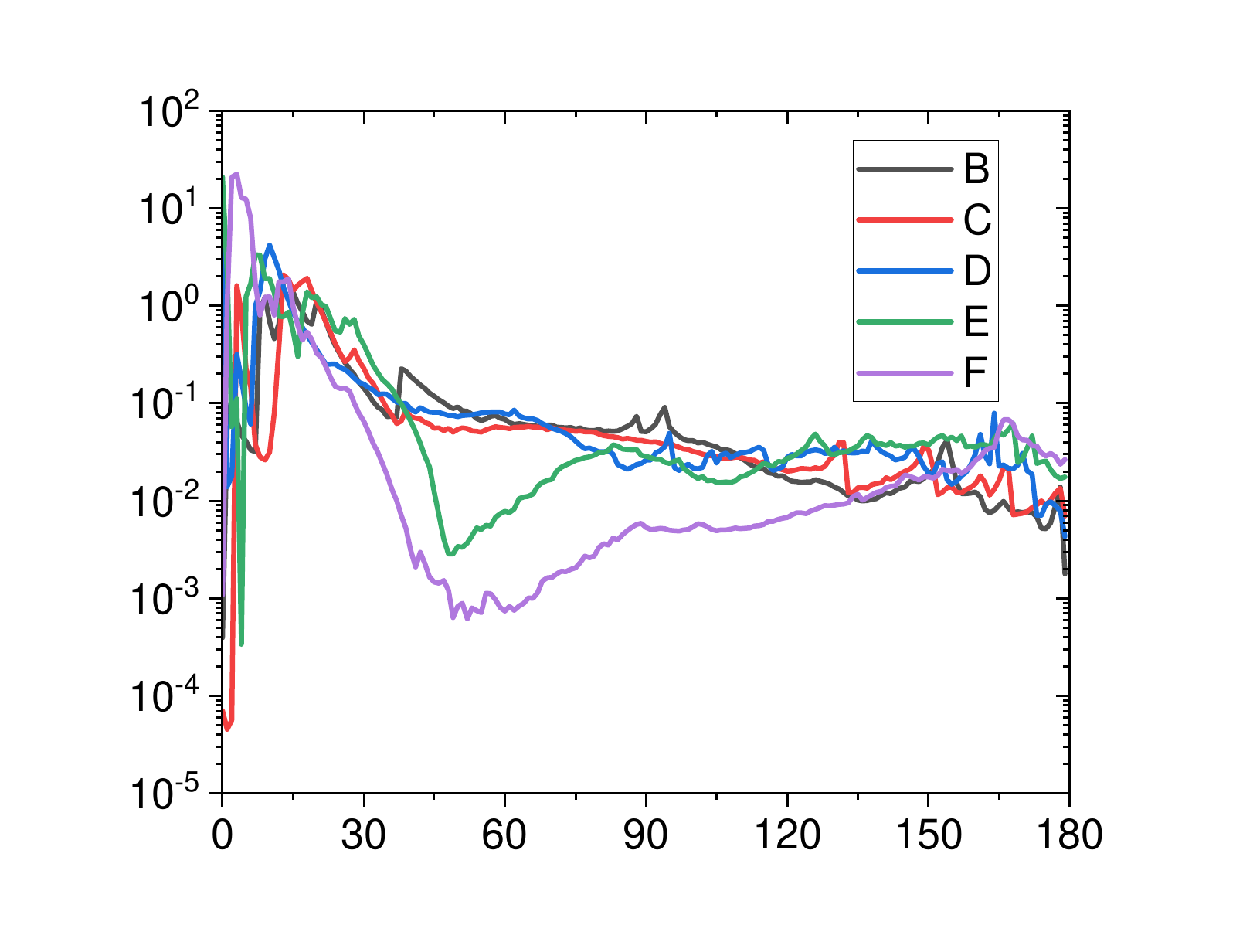}
  \put(20,15){\small $M_{11}$}
\end{overpic}
\begin{overpic}[width=0.48\textwidth]{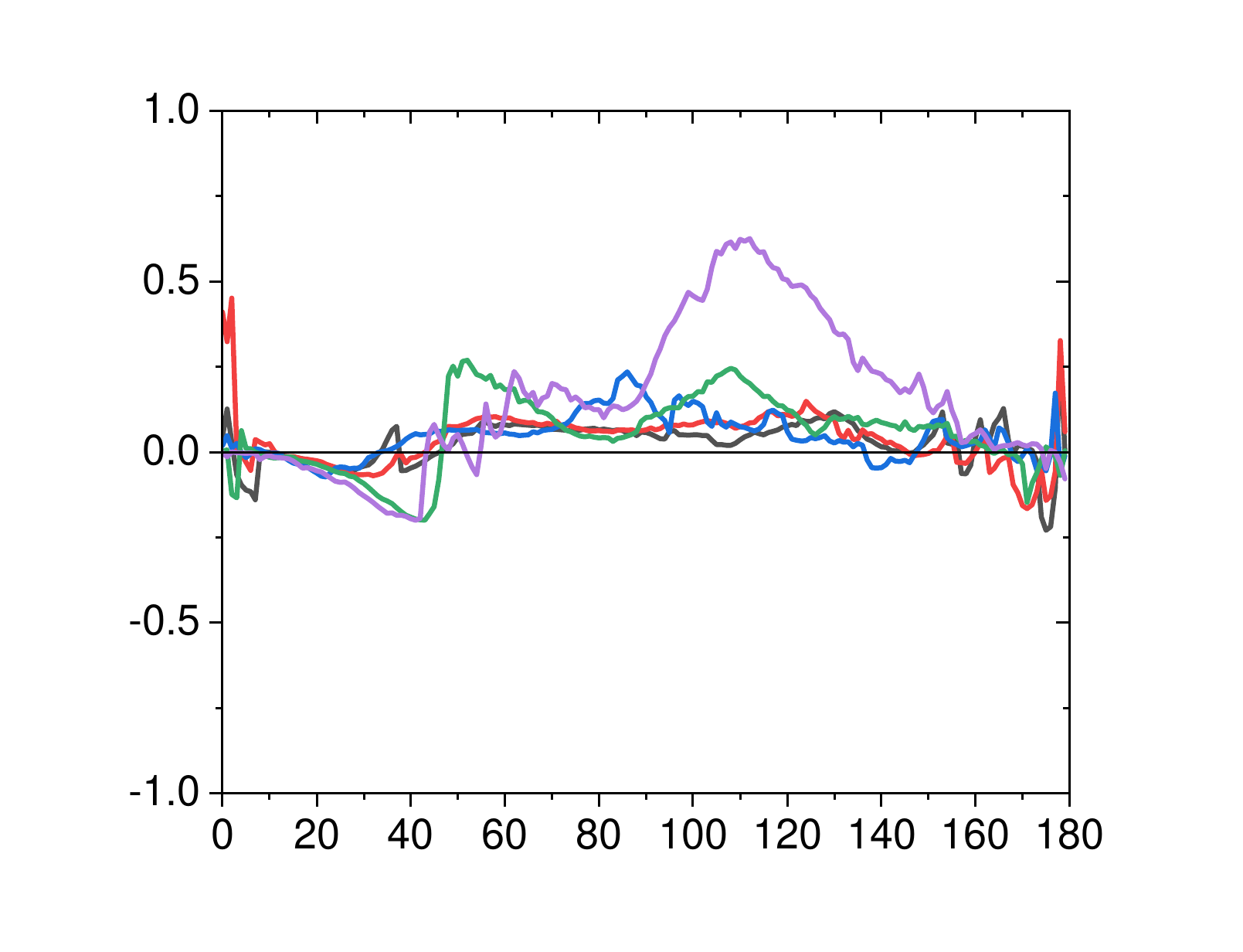}
  \put(20,15){\small $M_{12}/M_{11}$}
\end{overpic}
%---------------------------------------------
\begin{overpic}[width=0.48\textwidth]{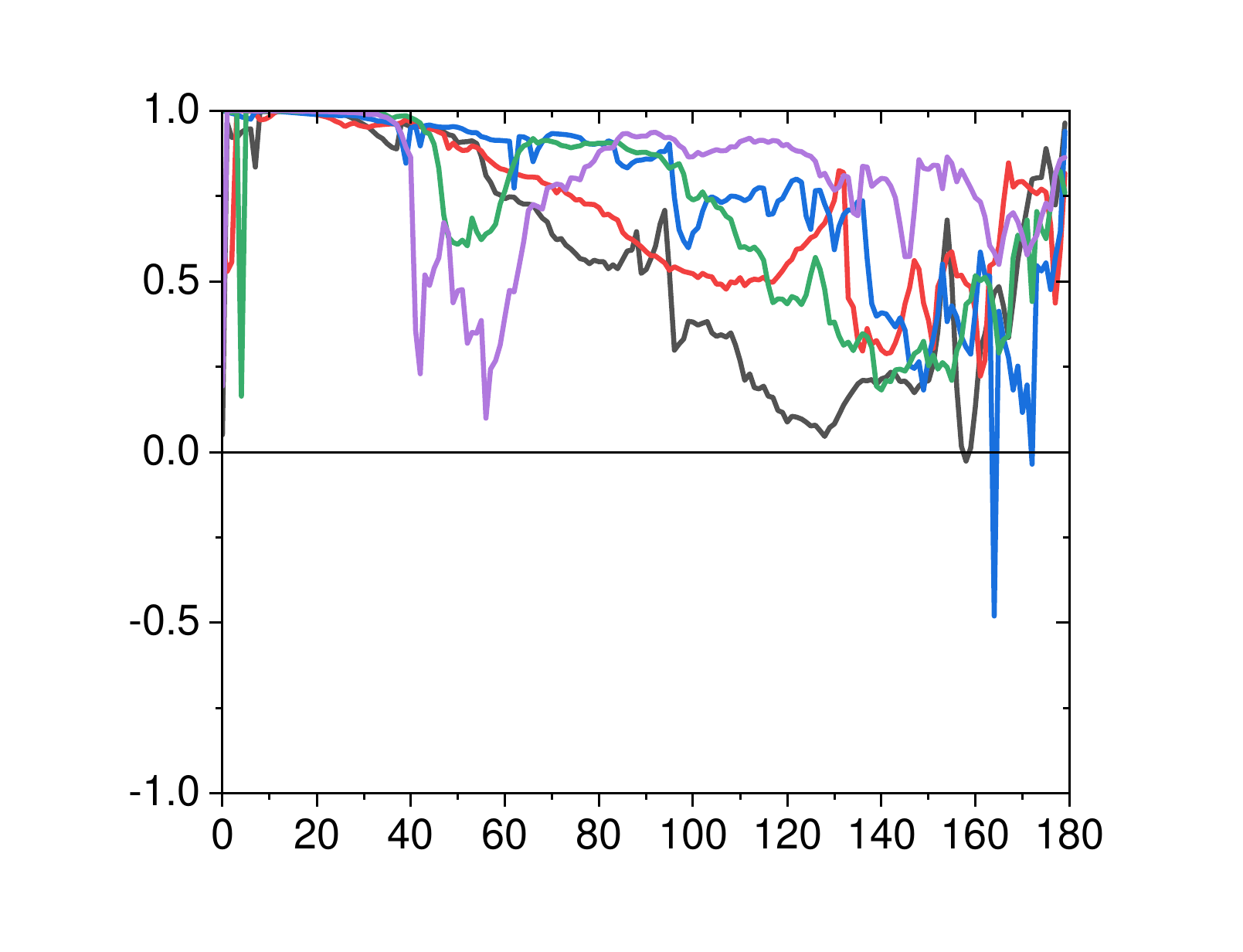}
  \put(20,15){\small $M_{22}/M_{11}$}
\end{overpic}
\begin{overpic}[width=0.48\textwidth]{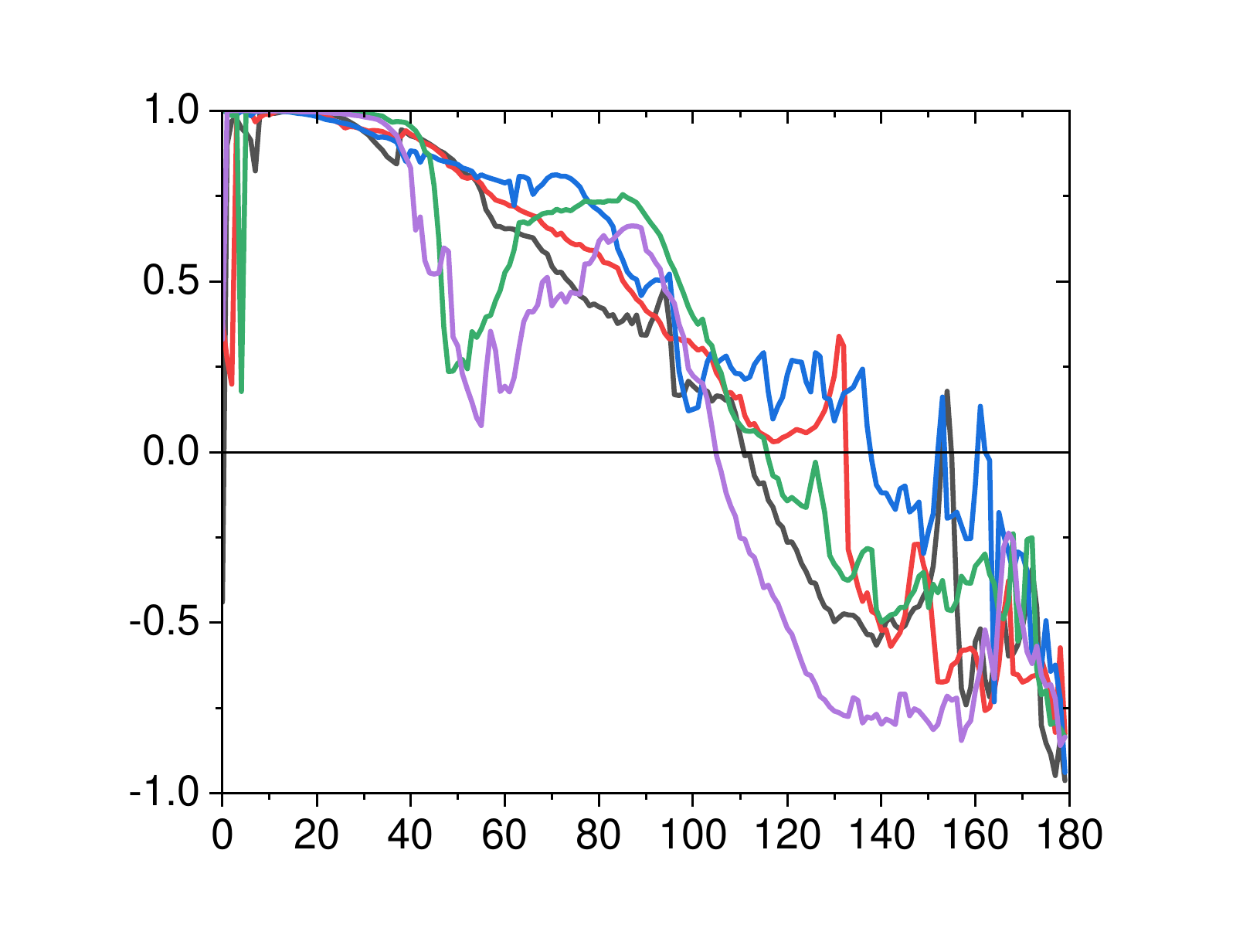}
  \put(20,15){\small $M_{33}/M_{11}$}
\end{overpic}
%---------------------------------------------
\begin{overpic}[width=0.48\textwidth]{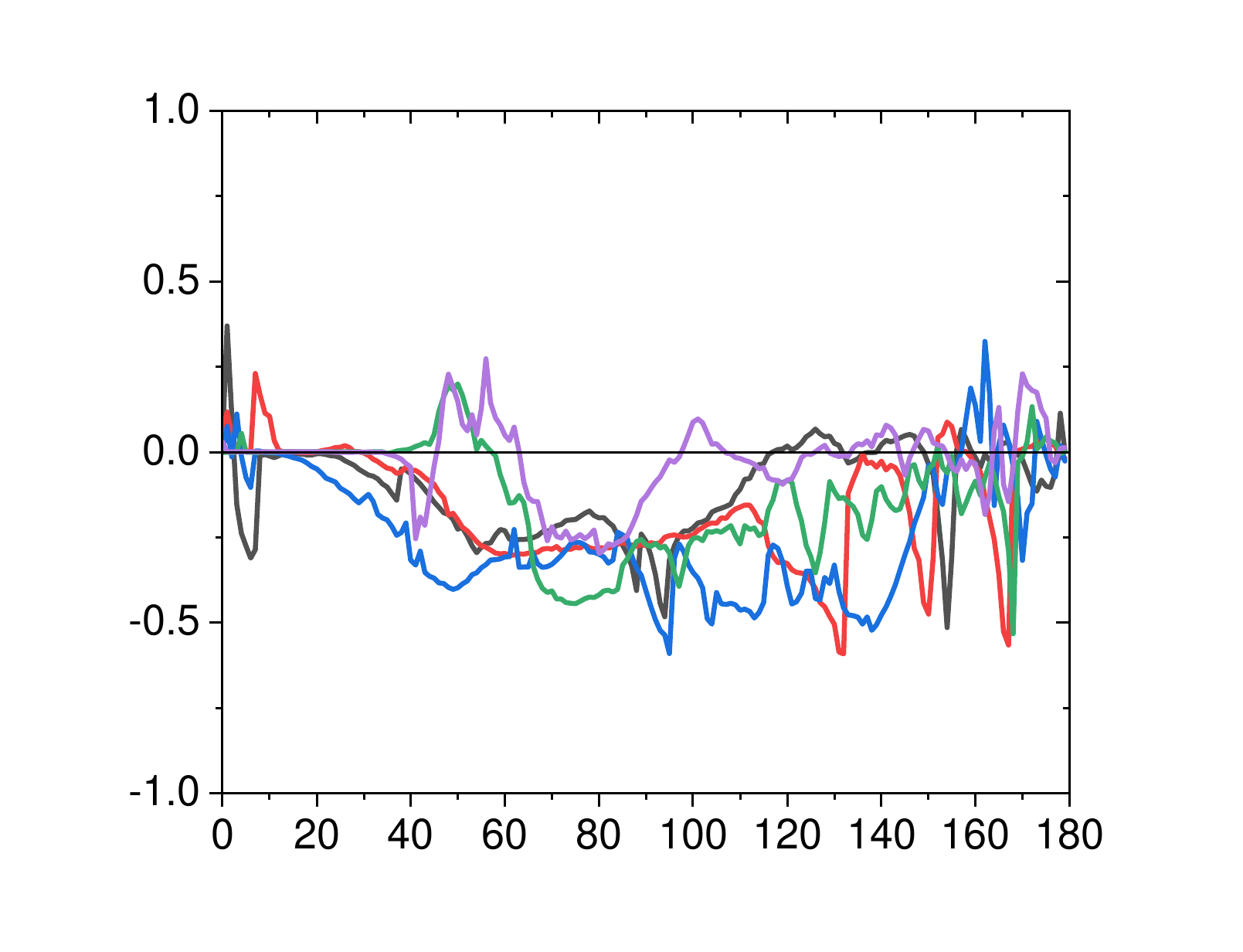}
  \put(20,15){\small $M_{34}/M_{11}$}
\end{overpic}
\begin{overpic}[width=0.48\textwidth]{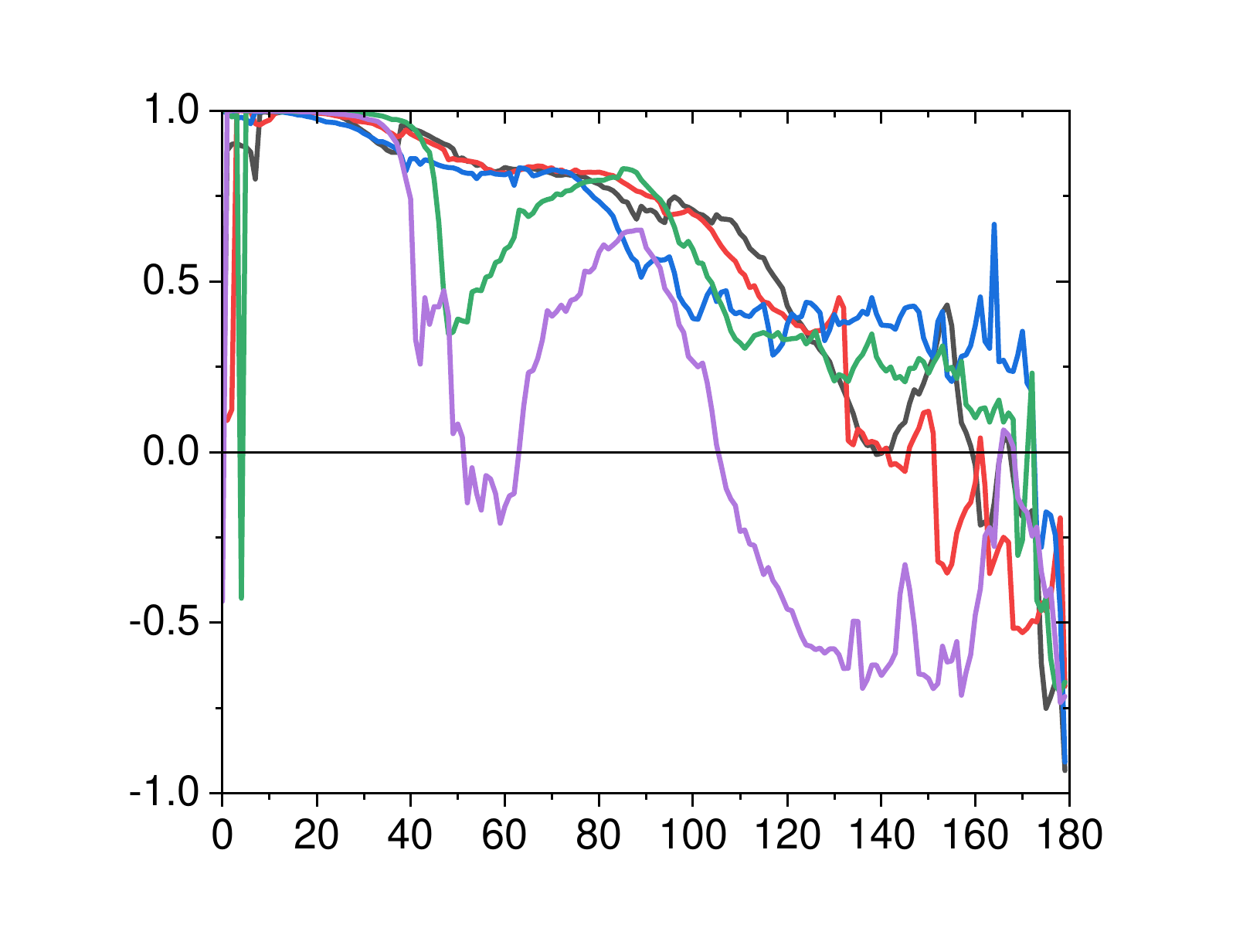}
  \put(20,15){\small $M_{44}/M_{11}$}
\end{overpic}

\caption{Scattering matrix elements for 5 individual randomly oriented convex hulls with NIP$ = 6, 8, 13, 18, 30$, corresponding to curves B, C, D, E, F in Figure~\ref{Fig4} respectively. The horizontal axis represents the scattering angle (in degrees).}
\label{Fig5}
\end{figure*}

To investigate how the geometrical irregularity of ice crystal shapes affects their single scattering properties, the scattering matrices were calculated for five randomly generated convex hulls with different Numbers of Initial Points (NIP = 6, 8, 13, 18, 30). To generate each convex hull, a set of points was randomly and uniformly sampled within the cube $[-1,1]^3$. The resulting convex hulls are shown in panels B--F of Figure~\ref{Fig4}, respectively. The number of traced rays is set to $10$ for each orientation, and the number of sampled orientation is set to $10^6$, the maximum number of internal reflections inside the scatterer is set to $5$. The calculated single scattering matrices are presented in Figure \ref{Fig5}.

%%--------------------------------------------------------------Fig5

From Figure ~\ref{Fig5}, it can be seen that the scattering matrix elements of individual randomly irregular particles exhibit more small spike-like fluctuations compared with those of single hexagonal crystals shown in Figure ~\ref{Fig3}. In the $M_{11}$ panel of Figure ~\ref{Fig5}, the scattering near the forward direction no longer shows the typical delta transmission feature characteristic of hexagonal crystals, but instead displays several small dips. This behavior is also observed in other nonzero matrix elements, such as $M_{22}/M_{11}$, $M_{33}/M_{11}$, and $M_{44}/M_{11}$. From the panels $M_{22}/M_{11}$, $M_{33}/M_{11}$, and $M_{44}/M_{11}$, we also find that in the forward scattering region, approximately from $0^\circ$ to $40^\circ$, the original polarization state is generally not significantly altered. Starting from around $40^\circ$, depolarization effects begin to become apparent.

\begin{figure*}[htbp]
\centering

\begin{overpic}[width=0.48\textwidth]{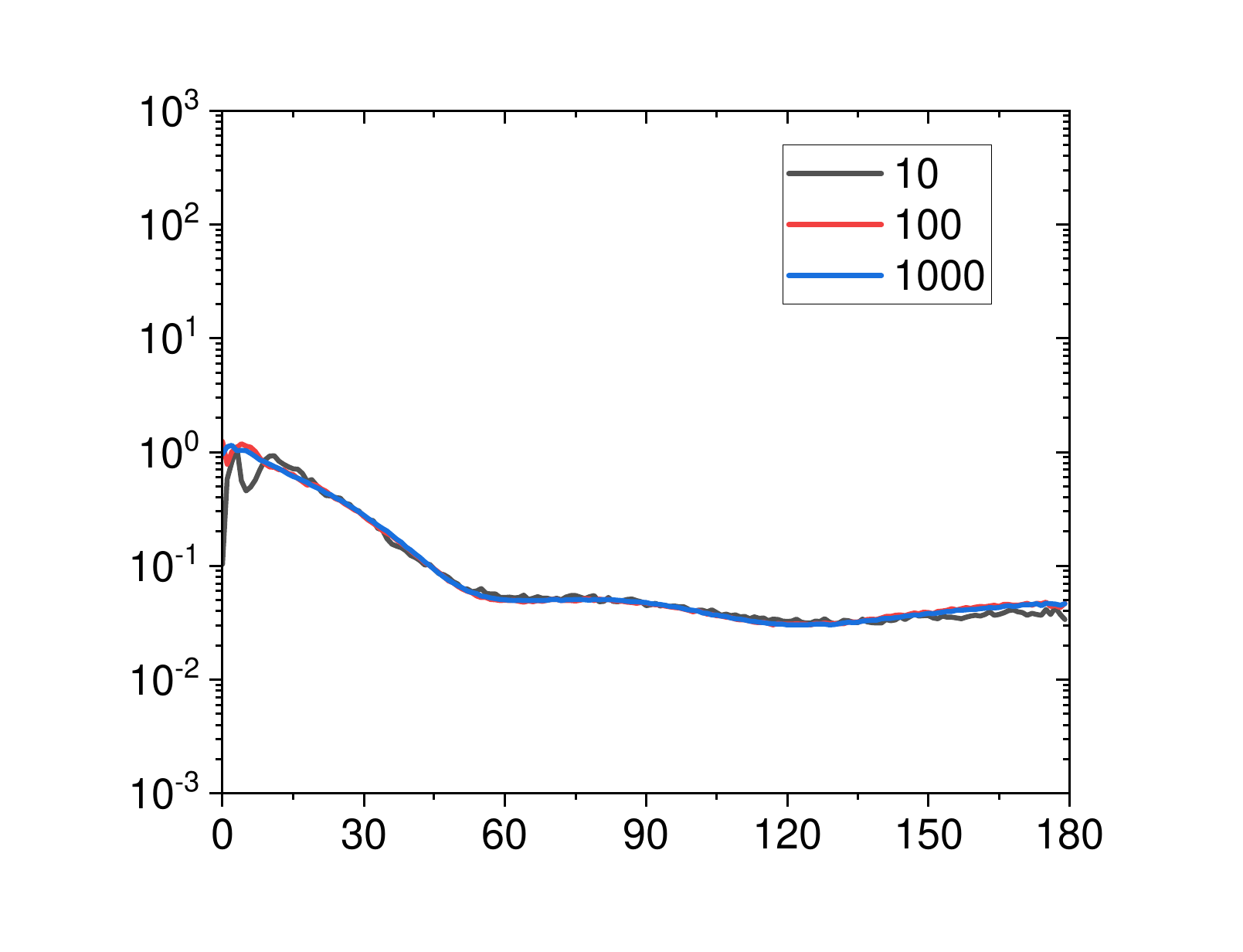}
  \put(20,15){\small $M_{11}$}
\end{overpic}
\begin{overpic}[width=0.48\textwidth]{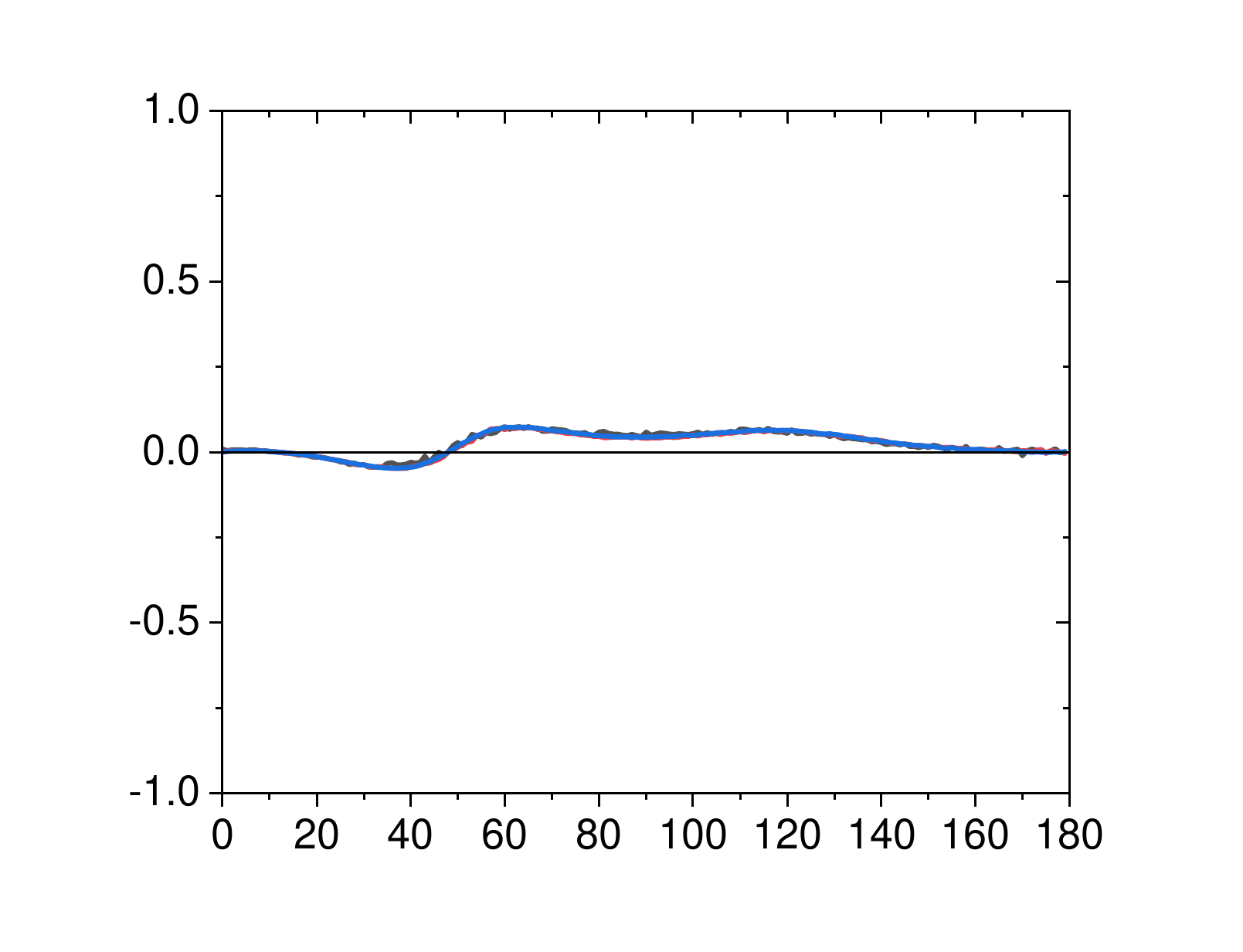}
  \put(20,15){\small $M_{12}/M_{11}$}
\end{overpic}
%---------------------------------------------
\begin{overpic}[width=0.48\textwidth]{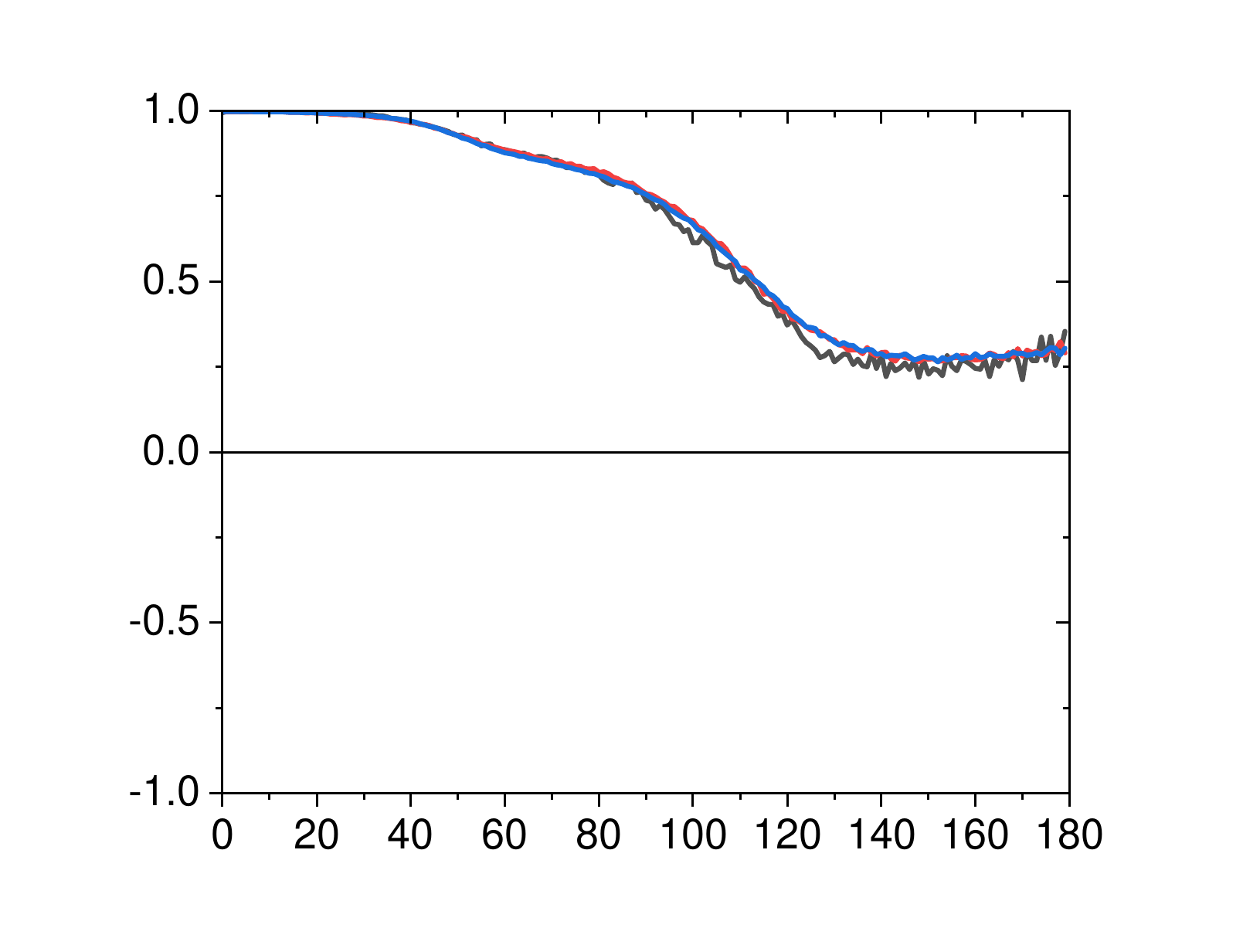}
  \put(20,15){\small $M_{22}/M_{11}$}
\end{overpic}
\begin{overpic}[width=0.48\textwidth]{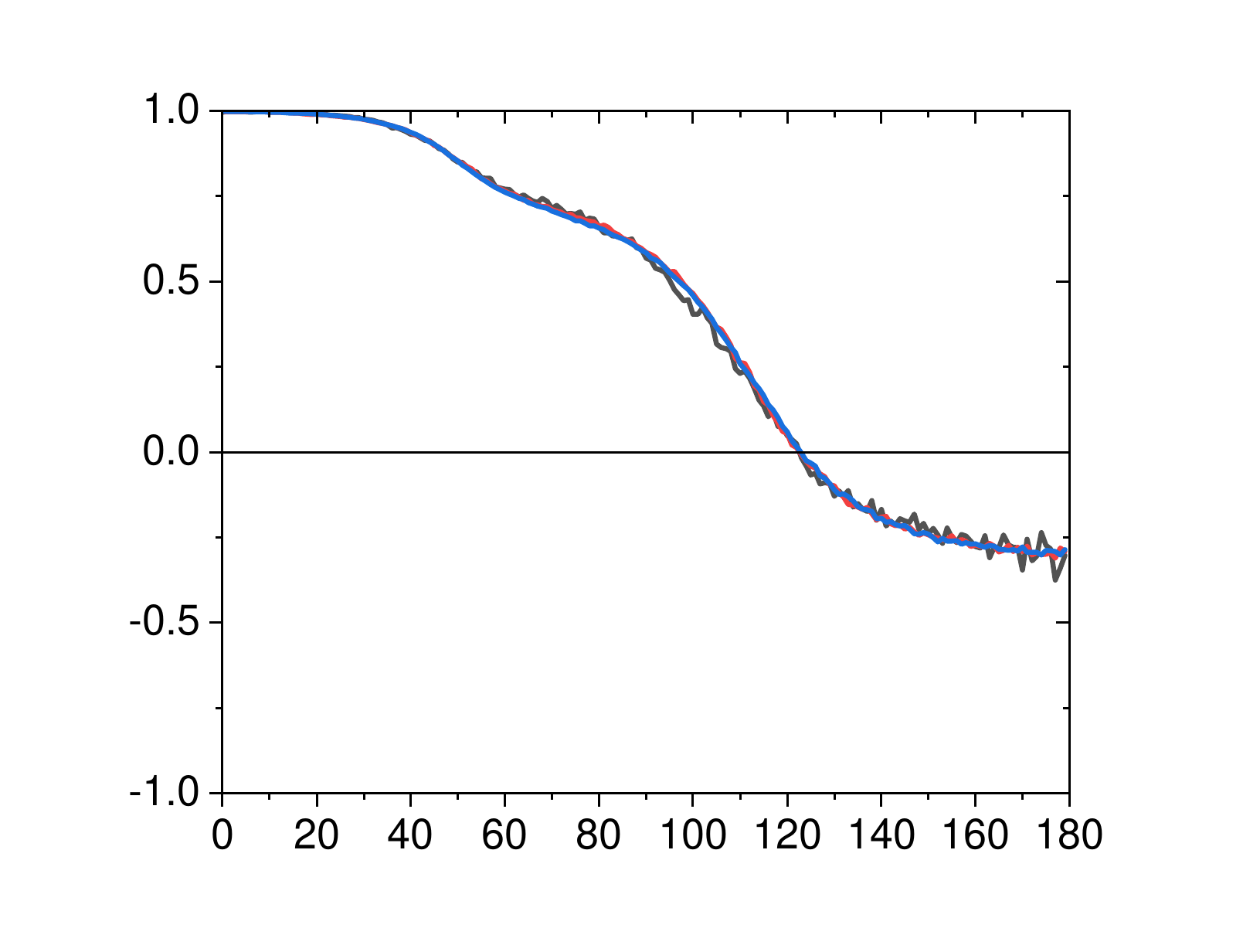}
  \put(20,15){\small $M_{33}/M_{11}$}
\end{overpic}
%---------------------------------------------
\begin{overpic}[width=0.48\textwidth]{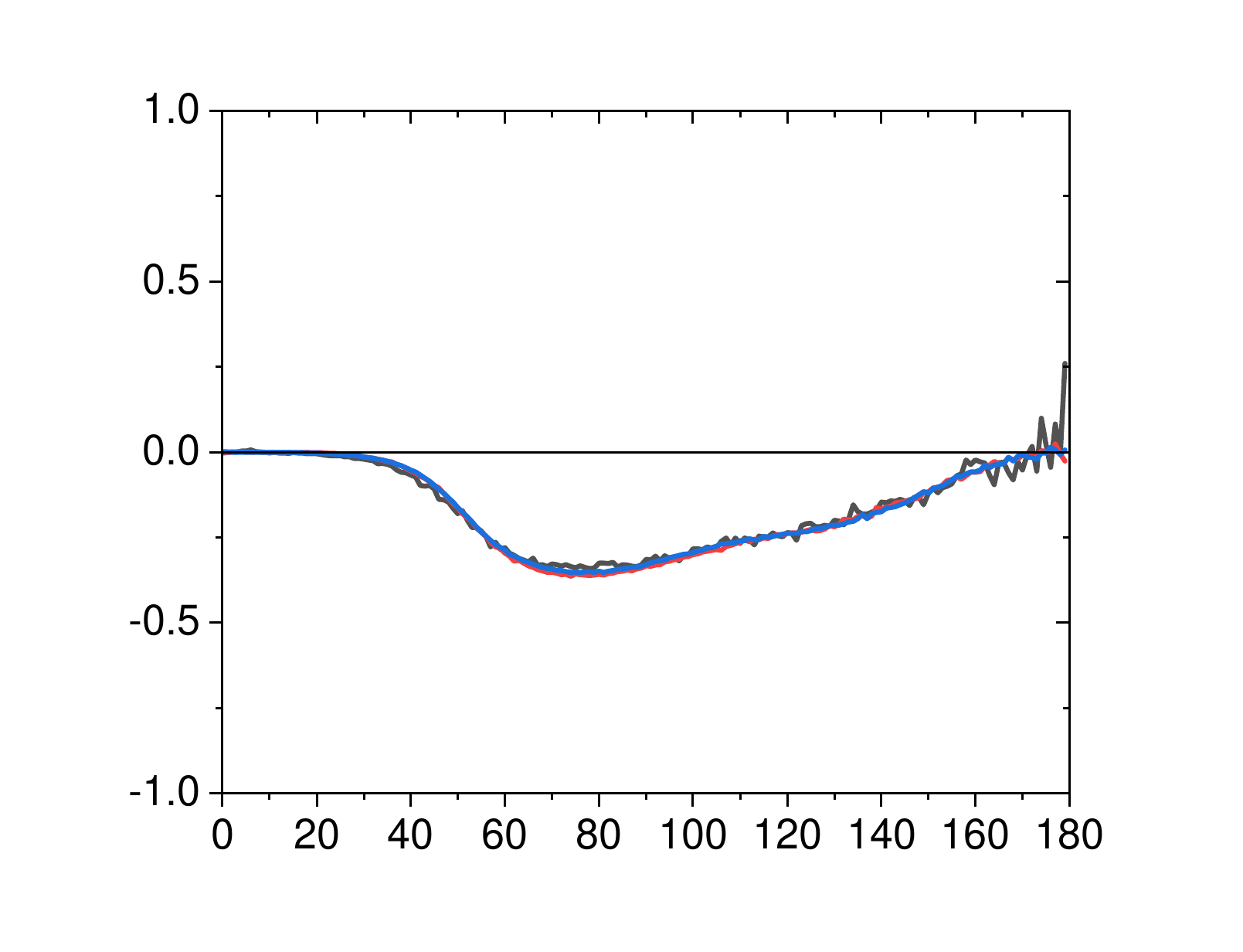}
  \put(20,15){\small $M_{34}/M_{11}$}
\end{overpic}
\begin{overpic}[width=0.48\textwidth]{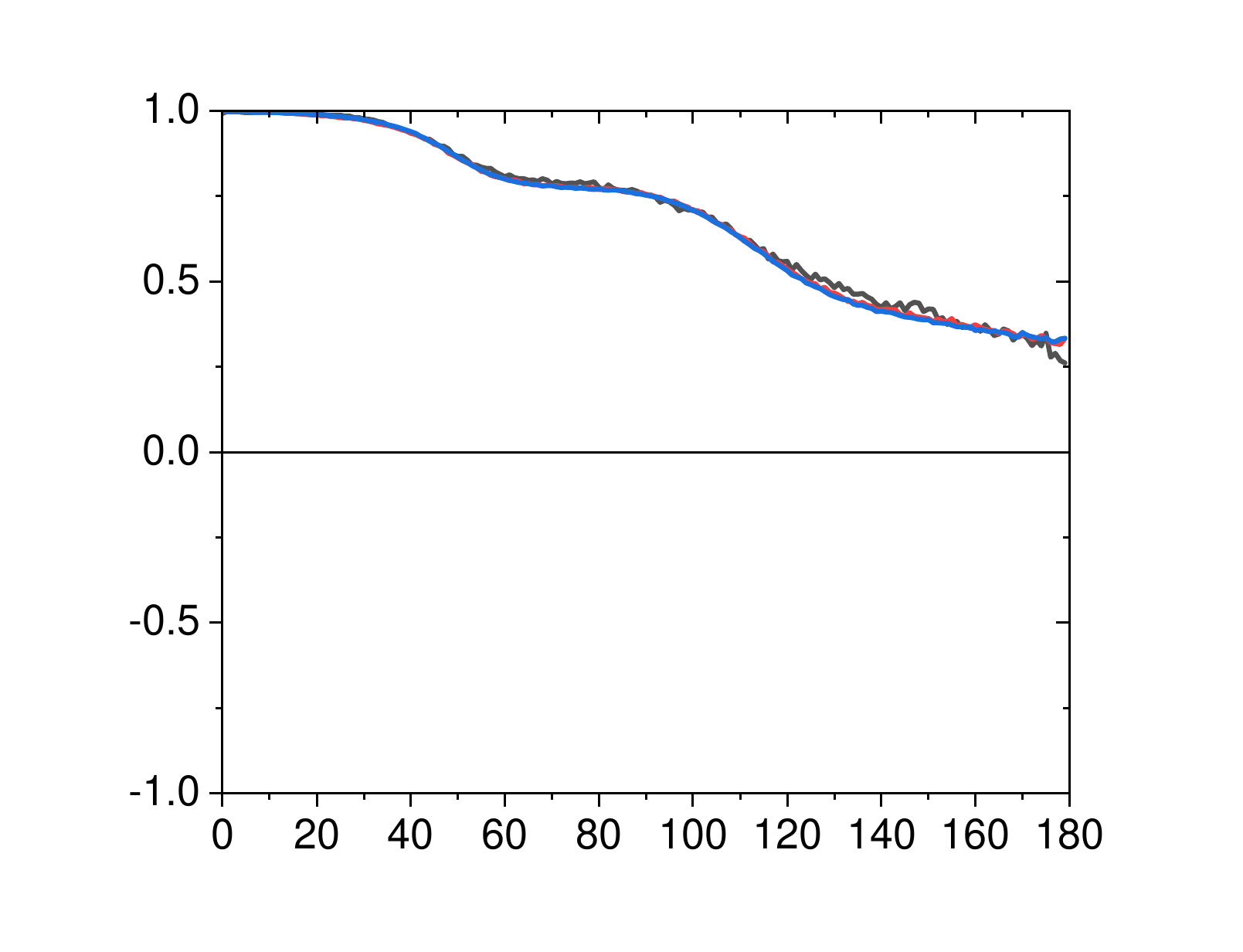}
  \put(20,15){\small $M_{44}/M_{11}$}
\end{overpic}

\caption{Single-scattering matrix elements computed for three ensembles of convex hulls, each consisting of 10, 100, and 1000 randomly generated particles. The horizontal axis represents the scattering angle (in degrees).}
\label{Fig7}
\end{figure*}

In Figure~\ref{Fig7}, the six scattering matrix elements are shown for three ensembles of random convex hulls, containing 10, 100, and 1000 particles, respectively. For each ensemble, the number of initial points NIP is uniformly sampled from intervals $[4,30]$. The number of traced rays per orientation, the number of sampled orientations, and the maximum number of internal reflections within the scatterer are set the same as in the numerical experiments used to calculate the scattering matrices of hexagonal prism ensembles.

%%--------------------------------------------------------------Fig7

From Figure~\ref{Fig7}, for ensembles containing different numbers of randomly shaped irregular particles, all non-zero scattering matrix elements exhibit much smoother curves without characteristic sharp peaks. This indicates that the smoothing effect is not limited to the $M_{11}$ element; other polarization-related elements also show an overall smooth behavior. Moreover, the single-scattering matrix elements obtained from statistical simulations of 100 and 1000 particles of different random shapes are highly consistent across the entire scattering angle range from $0^\circ$ to $180^\circ$. This implies that further increasing the number of particles would not significantly affect the scattering features. The $M_{11}$ in Figure~\ref{Fig7} agrees well with our calculations under the scalar model \cite{kargin2022numerical}, and is also consistent with the findings in \cite{liu2014effective}, which showed that the roughening of the surface and irregularization of facial geometry have similar influences on the scattering phase matrices.

By analyzing $M_{11}$, $M_{22}/M_{11}$, $M_{33}/M_{11}$, and $M_{44}/M_{11}$, we can see that the single-scattering intensity of ensemble of different randomly shaped particles is primarily concentrated in the forward scattering range of $0^\circ\!-\!40^\circ$ (note that diffraction is not included yet). Within this range, the polarization state, including linear polarization along the $x,y$ and $\pm 45^\circ$ directions, as well as circular polarization, remains almost unchanged. Beyond approximately $40^\circ$, however, the polarization state begins to exhibit noticeable changes. This is consistent with our calculated scattering matrix results for individual ice crystals (Figure~\ref{Fig5}).

By comparing Figures~\ref{Fig3} and ~\ref{Fig5}, and Figures~\ref{Fig6} and ~\ref{Fig7}, we can conclude that scattering matrix elements allow us to distinguish between regular and irregular particles. In general, the scattering matrix elements of an individual regular particle are relatively smooth, whereas those of an individual irregular particle often exhibit small local fluctuations. The single-scattering matrix of an ensemble of regular particles tends to show distinct features, while that of an ensemble of irregular particles generally appears smooth and featureless. In practice, many observational data indicate that, in order to accurately reproduce the measured scattering features, ensembles composed of particles with different sizes and shapes may be required, and even both regular and irregular particles need to be included in certain proportions (see, for example, work \cite{chepfer2001ice,BARAN20091239}, Fig.5.25 in book \cite{liou2016light}).

\section{Conclusions}
\label{section 4}

In this study, using the unified computational framework for scattering matrices of convex polyhedra proposed in our previous work \cite{Mu2026}, we conducted a series of statistical numerical experiments on single-scattering matrices for both regular and irregular particles. The framework enables flexible modification of ice crystal geometric shapes and was applied throughout the computations. Based on the results, we summarize the following main conclusions:

1. The non-zero elements of the scattering matrices of regular and irregular particles differ significantly, indicating that one can distinguish between regular and irregular particles based on their scattering matrices.

2. For ensembles composed of irregular particles, the non-zero elements of the single-scattering matrix, obtained in a statistical sense, are smooth and featureless, including those related to polarization. 

3. The scattering matrix elements of ensembles of hexagonal regular particles with various aspect ratios still reflect the common features of this particle type, such as the fact that the angles between any two faces remain unchanged.

4. The scattering characteristics of particle ensembles converge as the number of particles increases; beyond a certain ensemble size, further increases in particle number do not lead to significant changes in the overall scattering features.

5. Hexagonal extremely elongated columnar crystals and extremely thin plate-like crystals do not exhibit scattering matrix elements that act as bounding curves for particles with intermediate aspect ratios over all scattering directions.

6. No linear correlation is observed between the aspect ratio of hexagonal crystals and the values of the scattering matrix elements over all scattering directions. Such a relationship is only found to be possible for certain scattering matrix elements in the vicinity of specific scattering angles.

7. Columnar and plate-like hexagonal crystals can be distinguished from each other using specific scattering matrix elements at selected scattering angles.

It should be noted that the above conclusions are obtained under single-wavelength conditions and neglect absorption and diffraction effects. Moreover, due to the limited number of numerical experiments performed, further investigation is required to assess the robustness of these conclusions. In future work, diffraction and absorption effects will be incorporated into the unified computational framework to further enhance its physical realism and extend its applicability to a wider range of scattering scenarios. The results presented in this paper provide a clearer understanding of the influence of large ice crystal geometries on scattering properties and of how ensemble scattering characteristics emerge from particles of varying shapes. These findings can inform the interpretation of remote sensing data and support the improvement of atmospheric models.

\section*{Acknowledgment}
This work was partially supported by the Shenzhen Sci-Tech Fund (RCJC20231211090030059), National Key Research and Development Program of China (2025YFE0113-400) and National Natural Science Foundation of China (W2421102).

\bibliographystyle{unsrt}
\bibliography{2025Paper02}

%\end{sloppypar}
\end{document}